%% file: main.tex
\documentclass{article}

\usepackage{microtype}
\usepackage{graphicx}
\usepackage{subfigure}
\usepackage{booktabs} 

\usepackage{hyperref}

\usepackage[accepted]{icml2025}

\usepackage{amsmath}
\usepackage{amssymb}
\usepackage{mathtools}
\usepackage{amsthm}

\input{includes/base}
\input{includes/commands}
\input{includes/symbols}

\usepackage[capitalize,noabbrev]{cleveref}

\theoremstyle{plain}
\theoremstyle{definition}
\theoremstyle{remark}

\usepackage{todonotes} 
\usepackage{color-edits} 
\addauthor[Hai]{hvh}{blue}

\newcounter{phase}[algorithm]
\newlength{\phaserulewidth}
\newcommand{\setphaserulewidth}{\setlength{\phaserulewidth}}

\setphaserulewidth{.7pt}

\begin{document}
\newcommand{\phase}[1]{%
  %
  \STATE\leavevmode\llap{\rule{\dimexpr\labelwidth+\labelsep}{\phaserulewidth}}\rule{\linewidth}{\phaserulewidth}
  \vspace{-3ex}
  \STATE\strut\refstepcounter{phase}\textit{Stage~\thephase~--~#1}
  \vspace{-1.25ex}\STATE\leavevmode\llap{\rule{\dimexpr\labelwidth+\labelsep}{\phaserulewidth}}\rule{\linewidth}{\phaserulewidth}}
  
\twocolumn[
\icmltitle{Efficient Image Restoration via Latent Consistency Flow Matching}

\icmlsetsymbol{equal}{*}

\begin{icmlauthorlist}
\icmlauthor{Elad Cohen}{yyy}
\icmlauthor{Idan Achituve}{yyy}
\icmlauthor{Idit Diamant}{yyy}
\icmlauthor{Arnon Netzer}{yyy}
\icmlauthor{Hai Victor Habi}{yyy}
\\
{\normalfont
{\{elad.cohen02,idan.achituve,idit.diamant,arnon.netzer,hai.habi\}@sony.com}}{}

\end{icmlauthorlist}
\icmlaffiliation{yyy}{Sony Semiconductor Israel, Israel}
\icmlcorrespondingauthor{Hai Victor Habi}{hai.habi@sony.com}
\vskip 0.3in
]

\printAffiliationsAndNotice{}

\begin{abstract}
Recent advances in generative image restoration (IR) have demonstrated impressive results. However, these methods are hindered by their substantial size and computational demands, rendering them unsuitable for deployment on edge devices. This work introduces ELIR, an Efficient Latent Image Restoration method. ELIR addresses the distortion-perception trade-off within the latent space and produces high-quality images using a latent consistency flow-based model. In addition, ELIR introduces an efficient and lightweight architecture. Consequently, ELIR is 4$\times$ smaller and faster than state-of-the-art diffusion and flow-based approaches for blind face restoration, enabling a deployment on resource-constrained devices. Comprehensive evaluations of various image restoration tasks and datasets show that ELIR achieves competitive performance compared to state-of-the-art methods, effectively balancing distortion and perceptual quality metrics while significantly reducing model size and computational cost. The code is available at: \textcolor{blue}{\url{https://github.com/eladc-git/ELIR}}.
\end{abstract}
\input{files/introduction}

\begin{figure*}
\centering
\hspace*{0.0cm}\includegraphics[scale=0.4]{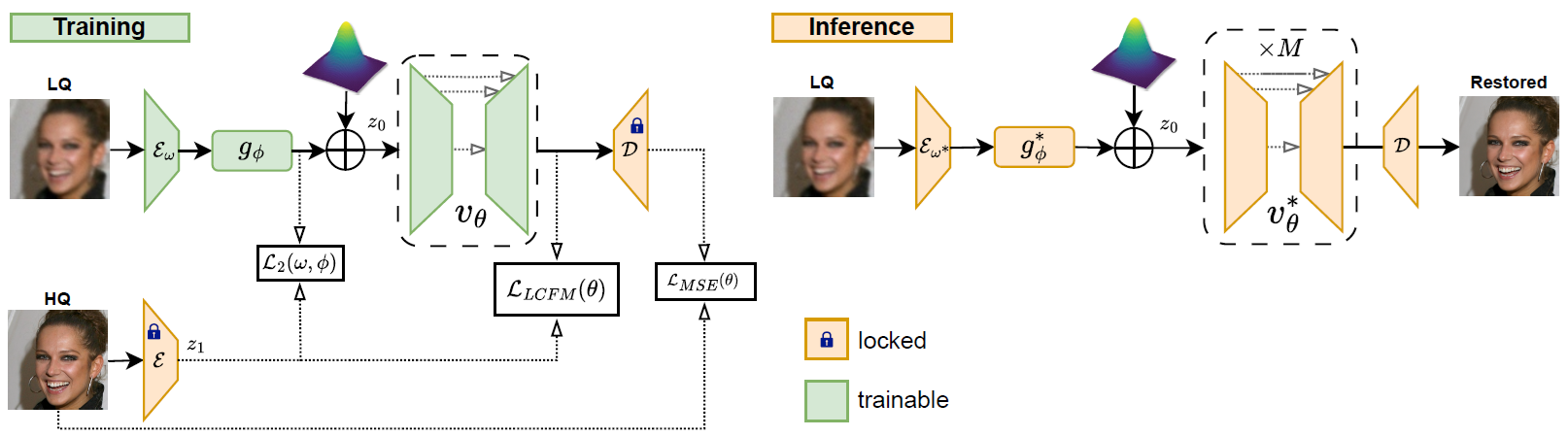}
\caption{\textbf{\name{} Overview.} During training, we optimize the encoder $\mathcal{E}_{\omega}$, coarse estimator $g_{\phi}$, and the vector field $\vectorsym{v}_{\theta}$ for a specific IR task. During inference, we predict a consistent linear direction from LQ toward the HQ images, yielding high-quality results and balancing distortion and perception. Both training and inference are conducted in the latent space. } 
\label{fig:scheme}
\end{figure*}
\input{files/related_work}

\input{files/preliminaries}
\input{files/method}

\input{files/experiments}

\section{Conclusions}\label{sec:conclusions}
This study introduces Efficient Latent Image Restoration (ELIR), an efficient IR method aiming to address the distortion-perception trade-off within the latent space. \name{} consists of an initial stage of a coarse $\ell_2$ estimator, whose goal is to reduce errors of the LQ image, followed by latent consistency flow matching (LCFM). The LCFM is a combination of latent flow matching and consistency flow matching that enables a small number of NFEs and a reduction of the evaluation cost. In addition, we propose an efficient neural network architecture to significantly reduce computational complexity and model size. We have evaluated ELIR on several IR tasks and shown state-of-the-art performance in terms of model efficiency. In terms of distortion and perceptual quality, we have shown competitive performance with state-of-the-art methods. Such improvement enables efficient deployment on resource-constrained devices. We leave additional efficient architectures exploration, such as Swin Transformers and encoder-decoder latent space dimensionality, as well as exploring \name{} under real-world or out-of-distribution degradation conditions for future work.

\bibliography{example_paper}
\bibliographystyle{icml2025}

\newpage
\onecolumn
\input{files/appendix}


\end{document}

%% file: includes/base.tex
\usepackage[inkscapearea=page]{svg}

\usepackage{mathtools}
\usepackage{hyperref}
\usepackage{amssymb}
\usepackage{bm}
\usepackage{graphicx}
\usepackage{float}
\usepackage{amsmath}
\usepackage{braket}
\usepackage{tikz}
\usepackage{bbm}
\usepackage{multirow}
\usetikzlibrary{positioning, backgrounds, fit, shapes.arrows}
\usetikzlibrary{decorations.markings}
\usepackage{amsthm}
\usepackage{cuted}
\usepackage{enumitem}
\usepackage{caption}
\usepackage{subcaption}

\newtheorem{theorem}{Theorem}[section]

\usepackage[table]{xcolor}
\usepackage{booktabs}
\usepackage{multirow}
\usepackage{tcolorbox}
\usepackage{bbding}
\usepackage{amssymb}
\usepackage{pifont}

%% file: includes/commands.tex
\newcommand\MakeUppercaseGreek[1]{
  \begingroup
    \let\psi\Psi
    \let\omega\Omega
    \let\gamma\Gamma
    \MakeUppercase{#1}
  \endgroup}
\newcommand{\vectorsym}[1]{\bm{#1}}

\newcommand{\expectation}[2]{\mathbb{E}_{#2}\left[#1\right]}

\newcommand{\brackets}[1]{\left(#1\right)}

\newcommand{\norm}[1]{\left\lVert#1\right\rVert}

\newcommand{\xmark}{\ding{55}}
\newcommand{\cmark}{\ding{51}}

%% file: includes/symbols.tex
\newcommand{\p}[0]{\vectorsym{\theta}}
\newcommand{\y}[0]{\vectorsym{y}}

\newcommand{\z}[0]{\vectorsym{z}}
\newcommand{\x}[0]{\vectorsym{x}}

\newcommand{\name}{ELIR}

%% file: files/introduction.tex
\section{Introduction}

Image restoration (IR) is a challenging low-level computer vision task focused on generating visually appealing high-quality (HQ) images from low-quality (LQ) images (e.g., noisy, blurry). Image deblurring \cite{kupyn2019deblurgan,whang2022deblurring}, blind face restoration \cite{wang2021towards,li2020blind}, image super-resolution \cite{dong2012nonlocally,dong2015image}, denoising \cite{delbracio2023inversion} and inpainting \cite{yu2019free} can be categorized under IR. Algorithms that tackle the IR problem are commonly evaluated by two types of metrics: 1) a distortion metric (e.g. PSNR) that quantifies some type of discrepancy between the reconstructed images and the ground truth; 2) a perceptual quality metric (e.g. FID \cite{heusel2017gans}) that intends to assess the appeal of reconstructed images to a human observer. The distortion and perception metrics are usually at odds with each other, leading to a distortion-perception trade-off \cite{blau2018perception}. This trade-off can be viewed as an optimization problem of minimizing distortion while achieving a \emph{bounded} perception index \cite{freirich2021a}.

\begin{figure}
\centering
\includesvg[width=0.5\textwidth]{images/psnr_fid_fps_size_v5.svg}
\caption{\textbf{\name's Performance:} Comparison between \name{} and state-of-the-art baseline methods. \name{} is the smallest and fastest method while maintaining competitive results. Metrics such as LPIPS and \#Params, where smaller is better, are inverted and normalized for display. The results were obtained using the CelebA-Test dataset for blind face restoration.}
\label{fig:performance}
\end{figure}

Recently, several approaches have explored this direction \cite{yue2024difface,lin2023diffbir,rombach2022high,zhu2024flowie,10681246,ohayon2024posterior}. Although these methods achieve state-of-the-art results, deploying them on edge devices such as mobile phones or image sensors is challenging due to significant model size and computational requirements. The high demands stem from three main reasons: (i) the transformer-based architecture used by these methods, which incurs substantial computational cost and model size; (ii) state-of-the-art approaches based on diffusion or flow matching necessitate multiple neural function evaluations (NFE) during inference, posing difficulties for resource-constrained devices; (iii) many methods operate directly in pixel space, demanding high computational costs, particularly at high resolutions.

In this work, we address the challenge of providing an efficient algorithm for IR that exhibits significantly improved efficiency in terms of model size and computational cost while maintaining comparable performance. We present \name{}, an Efficient Latent Image Restoration method to address the distortion-perception trade-off in latent space. We propose latent consistency flow matching (LCFM), an integration of latent flow matching \cite{dao2023flow} and consistency flow matching \cite{yang2024consistencyfm}. To the best of our knowledge, this approach is presented here for the first time. In addition, we suggest replacing the transformer-based architecture with a convolution-based one that can be deployed on resource-constrained devices. \name{} overcomes the high demand requirements by (1) working in latent space; (2) utilizing LCFM to decrease the number of NFEs, and (3) using convolutions instead of transformers. Consequently, \name{} significantly reduces the computational costs associated with processing high-resolution images. We conducted a set of experiments to validate \name{} and highlight its benefits in terms of distortion, perceptual quality, model size, and latency. Specifically, we evaluate \name{} on blind face restoration, super-resolution, denoising, and inpainting. In all tasks, we demonstrate significant efficiency improvements compared to diffusion and flow-based methods. We exhibit the smallest model size and significantly increase frames per second (FPS) processing speed. \name{} achieves these improvements without sacrificing distortion or perceptual quality, remaining competitive with state-of-the-art approaches (Fig.~\ref{fig:performance}). Our contributions are summarized as follows:
\begin{itemize}
    \item We introduce latent consistency flow matching (LCFM), which integrates latent and consistency flow matching for reducing the number of NFEs.
    \item We introduce Efficient Latent Image Restoration (ELIR), an efficient method addressing the distortion-perception trade-off within the latent space.
    \item We perform experiments using several datasets on various image restoration tasks, including blind face restoration, super-resolution, denoising, and inpainting. The results show a significant reduction in model size and latency compared to state-of-the-art diffusion and flow-based methods while maintaining competitive performance.
\end{itemize}

%% file: files/related_work.tex
\section{Related Work}
 
Various approaches have been suggested for image restoration \cite{zhang2018learning,zhang2021designing,luo2020unfolding,liang2021swinir,zhou2022codeformer,lin2023diffbir,yue2024difface,zhu2024flowie,ohayon2024posterior}. In recent years, solutions for IR based on generative methods, including GANs \cite{goodfellow2014generative}, diffusion models \cite{song2021denoising} and flow matching \cite{lipman2023flow}, have emerged, yielding impressive results.

\textbf{GAN-based methods.} GAN-based techniques have been proposed to address image restoration. BSRGAN \cite{zhang2021designing} and Real-ESRGAN \cite{wang2021realesrgan} are GAN-based methods that use an effective degradation modeling process for blind super-resolution. GFPGAN \cite{wang2021gfpgan} and GPEN \cite{yang2021gan} proposed to leverage GAN priors for blind face restoration. GPEN suggested training a GAN network for high-quality face generation and then embedding it into a network as a decoder before blind face restoration. GFPGAN connected a degradation removal module and a pre-trained face GAN by direct latent code mapping. CodeFormer \cite{zhou2022codeformer} also uses GAN priors by learning a discrete codebook before using a vector-quantized autoencoder. Similarly, VQFR \cite{gu2022vqfr} uses a combination of vector quantization and parallel decoding, enabling efficient and effective restoration.

\textbf{Diffusion-based methods.}
DDRM \cite{kawar2022denoising}, DDNM \cite{wangzero}, and GDP \cite{fei2023generative} are diffusion-based methods that have superior generative capabilities compared to GAN-based methods by incorporating the powerful diffusion model as an additional prior. Under the assumption of known degradations, these methods can effectively restore images in a zero-shot manner. ResShift \cite{10681246} proposed an efficient diffusion model that facilitates the transitions between HQ and LQ images by shifting their residuals. SinSR \cite{wang2024sinsr} and OSEDiff \cite{wu2024one} introduced single-step diffusion models for super-resolution. Recently, several approaches have suggested two-stage pipeline algorithms. DifFace \cite{yue2024difface} suggested such a method for blind face restoration, performing sampling from a transition distribution followed by a diffusion process. DiffBIR \cite{lin2023diffbir} proposed to solve blind image restoration by first applying a restoration module for degradation removal and then generating the lost content using a latent diffusion model.

\textbf{Flow-based methods.}
Recently, FlowIE \cite{zhu2024flowie} and PMRF \cite{ohayon2024posterior} introduced two-stage algorithms for image restoration based on rectified flows \cite{liu2023flow}. FlowIE relies on the computationally intensive Stable Diffusion \cite{rombach2022high}, which limits its suitability for deployment on edge devices. PMRF has shown impressive results on both perception and distortion metrics by minimizing the MSE under a perfect perceptual index constraint. It alleviates the issues of solving the ODE by adding Gaussian noise to the posterior mean predictions. Nevertheless, PMRF uses sophisticated attention patterns that pose significant challenges for efficient execution on resource-constrained edge devices because of intensive shape and indexing operations \cite{li2022rethinking}. Our work introduces an efficient flow-based method designed with a hardware-friendly architecture, enabling its deployment on resource-constrained devices.

%% file: files/preliminaries.tex
\section{Preliminaries}
\begin{figure*}
\centering
\includegraphics[scale=0.095]{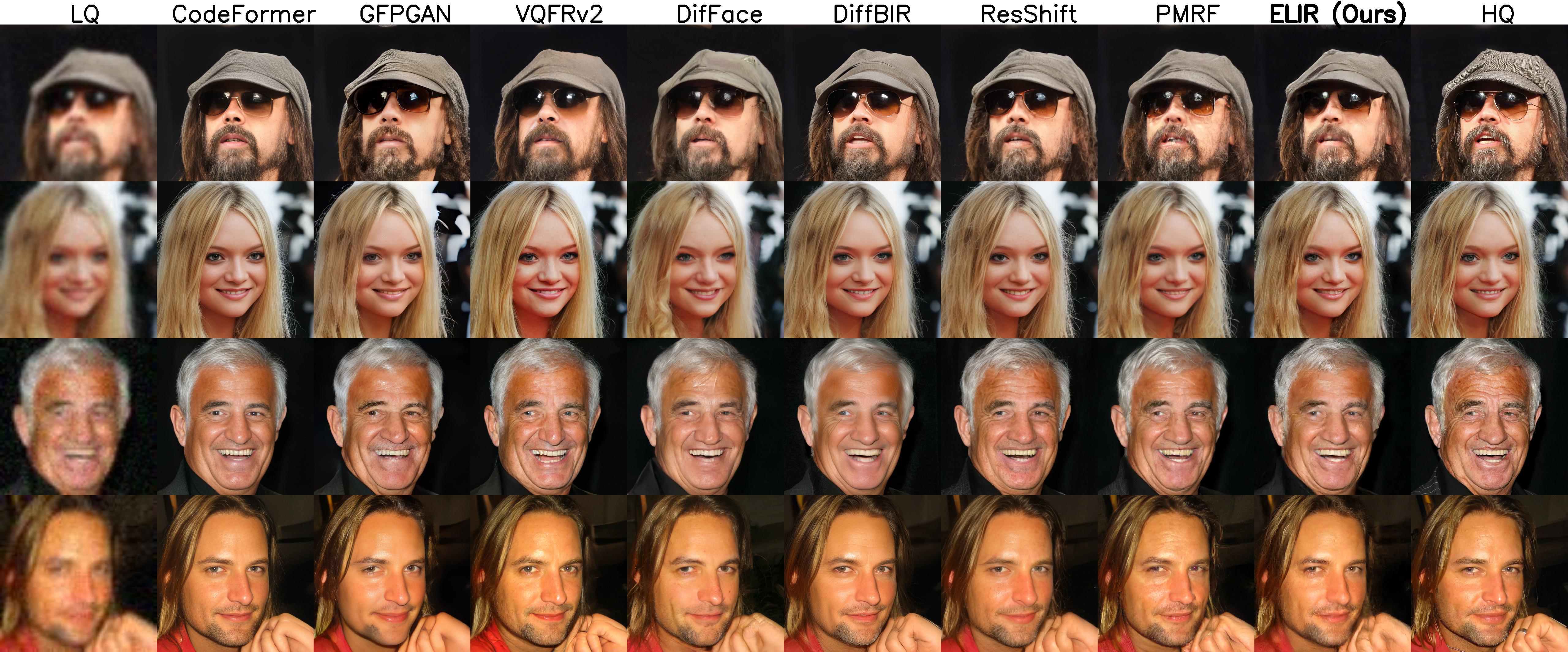}
\caption{\textbf{BFR Visual Results}. Visual comparisons between \name{} and baseline models sampled from CelebA-Test for blind face restoration. HQ and LQ refer to high-quality (ground truth) and low-quality (inputs) images.}
\label{fig:bfr}
\end{figure*}

\subsection{Distortion and Perception}
The perception of image quality is a complex interplay between objective metrics and subjective human judgment. While objective measures such as PSNR and SSIM are useful for quantifying distortion, they may not always correlate well with perceived image quality \cite{wang2004image}. Human observers are sensitive to artifacts and inconsistencies, even when they are subtle. Effective image restoration techniques must aim to minimize both objective distortion and perceptual artifacts, ensuring that the restored image is both visually pleasing and faithful to the original content. In this work, we denote the HQ and the corresponding LQ images as $\vectorsym{x}$ and $\vectorsym{y}$, respectively, and the reconstructed image by $\vectorsym{\hat{x}}$. Then, the distortion-perception trade-off can be formalized as \cite{freirich2021a}:
\begin{align}\label{distortion_perception}
    \min_{p_{\hat{\vectorsym{x}}|\vectorsym{y}}} \expectation{\norm{\vectorsym{x}-\hat{\vectorsym{x}}}_2^2}{} \quad s.t \quad W_2(p_{\hat{\vectorsym{x}}},p_{\vectorsym{x}}) \leq P,
\end{align}
where $W_2$ is the Wasserstein-2 distance, $p_{\vectorsym{x}}$ and $p_{\hat{\vectorsym{x}}}$ are the probability measures of the HQ and reconstructed image, respectively, and $P$ is the perception index.

\subsection{Consistency Flow Matching}
Consistency Flow Matching (CFM) \cite{yang2024consistencyfm} advances flow-based generative models \cite{chen2018neural,lipman2023flow,liu2023flow} by enforcing consistency among learned transformations. This constraint ensures that the transformations produce similar results regardless of the initial point. By utilizing ``straight flows'' for simplified transformations and employing a multi-segment training strategy, CFM achieves enhanced sample quality and inference efficiency. Specifically, given $\vectorsym{x}$ as an observation in the data space $\mathbb{R}^d$, sampled from an unknown data distribution, CFM first defines a vector field $\vectorsym{v}(\vectorsym{x}_t,t):\mathbb{R}^d\times[0,1]\xrightarrow{}\mathbb{R}^d$, that generates the trajectory $\vectorsym{x}_t \in \mathbb{R}^d$ through an ordinary differential equation (ODE): $\frac{d\vectorsym{x}_t}{dt} = \vectorsym{v}(\vectorsym{x}_t,t)$. \citet{yang2024consistencyfm} suggests training the vector field by the following velocity consistency loss:
\begin{align}\label{cfm_loss}
&\mathcal{L}_{CFM}\brackets{\vectorsym{\theta}}
=\\
&\mathbb{E}_{t}\expectation{\Delta f_{\vectorsym{\theta}}\brackets{\vectorsym{x}_t,\vectorsym{x}_{t+\Delta t},t}+\alpha\Delta v_{\vectorsym{\theta}}\brackets{\vectorsym{x}_t,\vectorsym{x}_{t+\Delta t},t} }{\vectorsym{x}_{t},\vectorsym{x}_{t+\Delta t}}\nonumber
\end{align}
where, 
\begin{align}
    &\Delta v_{\vectorsym{\theta}}\brackets{\vectorsym{x}_t,\vectorsym{x}_{t+\Delta t},t}=\norm{\vectorsym{v}_{\vectorsym{\theta}}(\vectorsym{x}_t,t)-\vectorsym{v}_{\vectorsym{\theta}^{-}}(\vectorsym{x}_{t+\Delta t},t+\Delta t)}_2^2,\nonumber\\
    &\Delta{f}_{\p}\brackets{\vectorsym{x}_t,\vectorsym{x}_{t+\Delta t},t}=\norm{f_{\vectorsym{\theta}}(\vectorsym{x}_t,t)-f_{\vectorsym{\theta}^{-}}(\vectorsym{x}_{t+\Delta t},t+\Delta t)}_2^2,\nonumber\\
    &f_{\vectorsym{\theta}}(\vectorsym{x}_t,t) = \vectorsym{x}_t + \brackets{1-t}\vectorsym{v}_{\vectorsym{\theta}}(\vectorsym{x}_{t},t).\nonumber
\end{align}
$t\sim\mathbb{U}[0,1-\Delta t]$ is the uniform distribution, $\Delta t$ is a small time interval and $\alpha$ is a positive scalar. $\theta^{-}$ denotes parameters without backpropogating gradients. The first term in \eqref{cfm_loss} ensures consistency in the end point using straight paths regardless of the location in the trajectory, while the first term matches the vector fields at all locations. To apply \eqref{cfm_loss}, we need to select a trajectory $\x_t$. Several options exist in the literature \cite{ho2020denoising,lipman2023flow,liu2023flow}. In this work, we use the optimal-transport conditional flow matching as proposed by \citet{lipman2023flow}, which enhances both the sampling speed and training stability. This trajectory is defined as $\vectorsym{x}_t=t\vectorsym{x}_1+(1-(1-\sigma_{min})t)\vectorsym{x}_0$, where $\vectorsym{x}_0$ and $\vectorsym{x}_1$ are sampled from source and target distributions, respectively, and $\sigma_{min}$ is a hyperparameter. During inference, utilizing the forward Euler method to solve the ODE reduces the number of NFEs compared to traditional Flow Matching (FM) techniques.

%% file: files/method.tex
\section{Method}
\begin{figure*}
\centering
\includegraphics[scale=0.19]{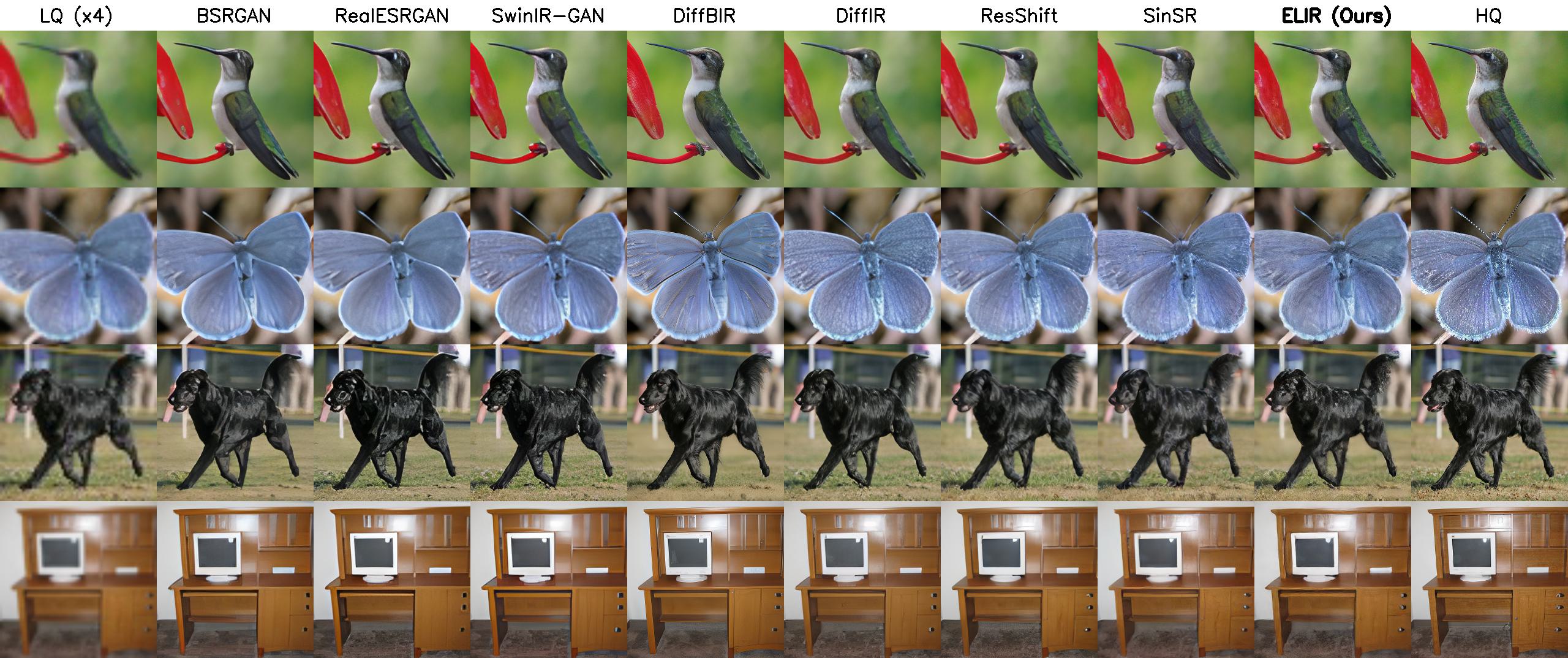}
\caption{\textbf{BSR Visual Results}. Visual comparisons between \name{} and baseline models sampled from ImageNet-Validation for blind super-resolution. HQ and LQ refer to high-quality (ground truth) and low-quality (inputs) images.}
\label{fig:bsr}
\end{figure*}
In this work, we address the challenge of developing an efficient method that minimizes average distortion under a bounded perceptual quality constraint as given in \eqref{distortion_perception}. Given an LQ image from an unknown degradation model, our goal is to restore it. The entire restoration process is performed in latent space, which enables efficient inference and significantly reduces the computational costs associated with processing high-resolution images. We denote $\mathcal{E}$ and $\mathcal{D}$ as the encoder and decoder trained on HQ images, respectively. As we show in the Appx.~\ref{apx:bound}, converging flow matching in latent space is influenced by both encoder-decoder and flow-matching optimization. The Wasserstein-2 distance between the HQ and reconstructed image distributions is bounded by $W_2(p_{\hat{\vectorsym{x}}},p_{\vectorsym{x}}) \leq \sqrt{\Delta_{\mathcal{E},\mathcal{D}}} + C\sqrt{\Delta_{\vectorsym{v}}}$, where $\Delta_{\mathcal{E},\mathcal{D}} $ and $\Delta_{\vectorsym{v}}$ are the encoder-decoder and latent flow matching objective errors, respectively, and $C$ is some constant. It demonstrates the crucial role of a well-designed encoder-decoder, as the bound depends on both the encoder-decoder error and the vector field error. We argue that this bound serves as the perception index $P$ from \eqref{distortion_perception}, justifying a latent space solution via flow matching. Here, we present \name{} (Efficient Latent IR), which aims to address the distortion-perception trade-off within the latent space. \name{} applies the Wasserstein-2 bound by leveraging an optimized pre-trained encoder-decoder, which minimizes the error term $\Delta_{\mathcal{E},\mathcal{D}}$ and by a latent consistency flow matching (LCFM) to reduce $\Delta_{\vectorsym{v}}$. We describe LCFM in Subsection~\ref{lcfm} and \name{}'s training, inference, and architecture in Subsection~\ref{elir}.
An overview of the proposed flow is presented in Fig.~\ref{fig:scheme}.

\subsection{Latent Consistency Flow Matching}\label{lcfm}
We introduce latent consistency flow matching (LCFM) as a combination of consistency flow matching \cite{yang2024consistencyfm} and latent flow matching \cite{dao2023flow}. LCFM approximates the transport between the latent representation of the source and target distributions. To achieve this, we define the latent representations of LQ and HQ images as $\vectorsym{z}_0$ and $\vectorsym{z}_1$, respectively. The optimal transport conditional flow from source to target distribution, as suggested by \citet{lipman2023flow}, is given by $\vectorsym{z}_t=t\vectorsym{z}_{1} + (1-(1-\sigma_{min})t)\vectorsym{z}_{0}$, where $t\in [0,1]$ is the time variable. To sample from the latent target distribution of $\vectorsym{z}_1$, we wish to obtain a vector field $\vectorsym{v}_{\theta}$ that would guide the direction of the linear path flowing from $\z_0$ to $\z_1$. To allow effective inference, we propose using multi-segment consistency loss \cite{yang2024consistencyfm} in the latent space. Specifically, given $K$ segments, the time interval $[0,1]$ is divided into $\set{[\frac{i}{K},\frac{i+1}{K}]}_{i=0}^{K-1}$. Then, the consistency loss of a segment is defined as:
\begin{align}\label{lcfm_loss}
&\mathcal{L}_{s}\brackets{\vectorsym{\theta},t}
=\\
&\expectation{\Delta f^{(i)}_{\vectorsym{\theta}}\brackets{\vectorsym{z}_t,\vectorsym{z}_{t+\Delta t},t}+\alpha\Delta v^{(i)}_{\vectorsym{\theta}}\brackets{\vectorsym{z}_t,\vectorsym{z}_{t+\Delta t},t} }{\vectorsym{z}_{t},\vectorsym{z}_{t+\Delta t}}\nonumber
\end{align}
where, 
\begin{align}
    &\Delta v_{\vectorsym{\theta}}^{(i)}\brackets{\vectorsym{z}_t,\vectorsym{z}_{t+\Delta t},t}=\norm{\vectorsym{v}^{(i)}_{\vectorsym{\theta}}(\vectorsym{z}_t,t)-\vectorsym{v}^{(i)}_{\vectorsym{\theta}^{-}}(\vectorsym{z}_{t+\Delta t},t+\Delta t)}_2^2,\nonumber\\
    &\Delta{f}^{(i)}_{\p}\brackets{\vectorsym{z}_t,\vectorsym{z}_{t+\Delta t},t}=\norm{f^{(i)}_{\vectorsym{\theta}}(\vectorsym{z}_t,t)-f^{(i)}_{\vectorsym{\theta}^{-}}(\vectorsym{z}_{t+\Delta t},t+\Delta t)}_2^2,\nonumber\\
    &f^{(i)}_{\vectorsym{\theta}}(\vectorsym{z}_t,t) = \vectorsym{z}_t + \brackets{\frac{i+1}{K}-t}\vectorsym{v}^{(i)}_{\vectorsym{\theta}}(\vectorsym{z}_{t},t).\nonumber
\end{align}
$t\sim\mathbb{U}[0,1-\Delta t]$ is the uniform distribution. Here, $i$ denotes the $i^{th}$ segment corresponding to time $t$, and $\Delta t$ and $\alpha$ are hyperparameters. $\vectorsym{v}^{(i)}_{\theta}(\vectorsym{z}_{t},t)$ is the vector field in the segment $i$ and $\theta^{-}$ denotes parameters without backpropogating gradients.

\subsection{Efficient Latent Image Restoration}\label{elir}
Here, we detail the training, inference, and architecture of \name{} as depicted in Fig~\ref{fig:scheme}.
\paragraph{Training.} Given optimized encode-decoder, we follow \cite{lin2023diffbir, zhu2024flowie} by applying a coarse $\ell_2$ estimator on the latent of the LQ input image, which plays a crucial role (see Appx.~\ref{apx:ablation}) in narrowing the probability direction path and is used as an initial point for the LCFM. Specifically, let $\mathcal{E}_{\omega}$ be a trainable encoder (parameterized by $\omega$) that projects an LQ image to the latent space, and $g_{\phi}$ be the coarse estimator (parameterized by $\phi$). The objective is to minimize the $\ell_2$ difference between the latent representations of the LQ and HQ images, which is given by $\mathcal{L}_{2}\brackets{\phi,\omega}= \expectation{\norm{\vectorsym{z}-\vectorsym{z}_1}_2^2}{\vectorsym{x},\vectorsym{y}}$, where $\vectorsym{z}=g_{\phi}(\mathcal{E}_{\omega}\brackets{\vectorsym{y}})$ and $\vectorsym{z}_1=\mathcal{E}(\vectorsym{x})$. During the optimization, $\mathcal{E}$ remains frozen, while $\mathcal{E}_{\omega}$ is trained in coordination with $g_{\phi}$. Since $\mathcal{E}_{\omega}$ is initialized with $\mathcal{E}$, its effectiveness may decrease when faced with unknown degradations such as denoising or inpainting unless it undergoes fine-tuning, as shown in the Appx.~\ref{apx:ablation}. Finetuning $\mathcal{E}_{\omega}$ improves the initial point of the flow, resulting in a reduction of $\Delta_{\vectorsym{v}}$. Next, we define $\vectorsym{z}_0=\vectorsym{z}+\vectorsym{\epsilon}$ as the source distribution samples with additive Gaussian noise $\vectorsym{\epsilon}$ having standard deviation $\sigma_{s}$. Adding such noise allows us to smooth the LQ embeddings density so it is well-defined over the entire space of HQ embeddings \cite{AlbergoV23}. Then, we utilize the latent consistency model from the source distribution of $\vectorsym{z}_0$ to a target distribution of $\vectorsym{z}_1$ using $\mathcal{L}_{LCFM}\brackets{\theta}$. Finally, we incorporate a mean square error (MSE) loss, detailed in \eqref{distortion_perception}, between our estimated outputs and the corresponding HQ images. The MSE loss is applied after continuing the latent representation $\vectorsym{f}^{(i)}_{\theta}(\vectorsym{z}_t,t)$ to $t=1$ by $\hat{\vectorsym{z}}_1=\vectorsym{f}^{(i)}_{\theta}(\vectorsym{z}_t,t) + \brackets{1-\frac{i+1}{K}}\vectorsym{v}_{\theta}\brackets{\vectorsym{f}^{(i)}_{\theta}(\vectorsym{z}_t,t),\frac{i+1}{K}}$, which produces the latent of the restored images. The loss is then calculated as $\mathcal{L}_{MSE}\brackets{\theta} = \expectation{\norm{\vectorsym{x}-\mathcal{D}\brackets{\hat{\vectorsym{z}}_1}}_2^2}{\vectorsym{x},\vectorsym{y}}$. Combining LCFM loss with the MSE loss, resulting in the following distortion-perception objective:
\begin{equation}
    \mathcal{L}_{DP}\brackets{\theta} = (1-\beta)\mathcal{L}_{LCFM}\brackets{\theta} + \beta\mathcal{L}_{MSE}\brackets{\theta},
\end{equation}
where $\beta$ is a hyperparameter for balancing perception and distortion
losses. Optimizing the total loss:
\begin{equation}
\mathcal{L} = \mathcal{L}_{2}\brackets{\phi,\omega}+\mathcal{L}_{DP}\brackets{\theta},
\end{equation}
which yeilds trained parameters $\omega^{*}$, $\phi^{*}$ and $\theta^{*}$. The losses $\mathcal{L}_{2}$ and $\mathcal{L}_{DP}$ are optimized jointly where the gradients of $\theta$ are detached from $\omega,\phi$. 

\paragraph{Inference.} During inference, we employ the trained encoder and coarse estimator by $\z=\mathcal{E}_{\omega^{*}}(g_{\phi^{*}}(\vectorsym{y}))$ and add a Gaussian noise with the same standard deviation $\sigma_{s}$. Then, we utilize the optimized vector field $\vectorsym{v}_{\theta^{*}}$ for solving the ODE using the forward Euler method with $M$ steps. Algorithm~\ref{alg} outlines the inference procedure.

\paragraph{Architecture.} We suggest an efficient and lightweight architecture consisting of only convolutional layers. We utilize Tiny AutoEncoder \cite{von-platen-etal-2022-diffusers}, a pre-trained tiny CNN version of Stable Diffusion VAE \cite{esser2024scaling}. It allows us to compress the image by a factor of 12 with only 1.2M parameters for each encoder and decoder. The coarse estimator consists of RRDB blocks \cite{wang2018esrgan} with 5.5M parameters, and we use U-Net \cite{ronneberger2015u} for the vector field. Additional details about the architecture can be found in the Appx.~\ref{apx:nn}.

\begin{algorithm}[t]
\caption{\name{} Algorithm}
\label{alg}
\begin{algorithmic}
\REQUIRE LQ image $\y$, number of Euler steps $M$, noise variance $\sigma_{s}^2$ 

\phase{Initial Point}
\STATE $\vectorsym{z} = g_{\phi^{*}}(\mathcal{E}_{\omega^{*}}(\vectorsym{y}))$
\STATE $\vectorsym{\epsilon}\sim\mathcal{N}(0,\sigma^2_{s}I)$
\STATE $\z_0 = \z + \vectorsym{\epsilon} $

\phase{Solve ODE (Euler Method)}
\STATE $\Delta=\frac{1}{M} $
\FOR{$i \gets 0,\Delta,...,1-\Delta$}
    \STATE $\hat{\z}_{i+\Delta} \gets \hat{\z}_{i} + \Delta \cdot \vectorsym{v}_{{\theta^{*}}}(\hat{\z}_{i},i)$ 
\ENDFOR
\STATE $\vectorsym{\hat{x}} = \mathcal{D}(\hat{\z}_1)$ \\
\RETURN $\vectorsym{\hat{x}}$
\end{algorithmic}
\end{algorithm}

%% file: files/experiments.tex
\begin{table*}[t]
\centering
\footnotesize

    \resizebox{\textwidth}{!}{%
    \begin{tabular}{lccccccccccc}
    \\\toprule
     & \multicolumn{2}{c}{} & \multicolumn{7}{c}{CelebA-Test} & \multicolumn{1}{l}{LFW} & \multicolumn{1}{l}{CelebAdult} \\ \cmidrule(l){4-10} \cmidrule(l){11-11} \cmidrule(l){12-12}
     & \multicolumn{2}{c}{\multirow{-2}{*}{Efficiency}} & \multicolumn{3}{c}{Perceptual Quality} & \multicolumn{4}{c}{Distortion} & \multicolumn{2}{c}{Perceptual Quality} \\ \cmidrule(l){2-3} \cmidrule(l){4-6} \cmidrule(l){7-10} \cmidrule(l){11-12}
    \multirow{-3}{*}{Model} & \#Params{[}M{]} $\downarrow$ & FPS $\uparrow$ & \multicolumn{1}{l}{FID $\downarrow$} & \multicolumn{1}{l}{NIQE $\downarrow$} & MUSIQ $\uparrow$ & PSNR $\uparrow$ & SSIM $\uparrow$ & LPIPS $\downarrow$ & IDS $\downarrow$ & FID $\downarrow$ & FID $\downarrow$ \\ \midrule\midrule
    CodeFormer  & 94 & {\color[HTML]{333333} 12.79} & 55.85 & 4.73 & {\color[HTML]{009901} \textbf{74.99}} & 25.21 & {\color[HTML]{009901} \textbf{0.6964}} & {\color[HTML]{FE0000} \textbf{0.3402}} & 37.41 & 53.46 & 115.42 \\
    GFPGAN  & {\color[HTML]{009901} \textbf{86}} & {\color[HTML]{FE0000} \textbf{26.37}} & 47.60 & 4.34 & {\color[HTML]{3531FF} \textbf{75.30}} & 24.98 & 0.6932 & 0.3627 & 36.14 & {\color[HTML]{009901} \textbf{49.51}} & 112.72 \\
    VQFRv2 & {\color[HTML]{3531FF} \textbf{83}} & {\color[HTML]{009901} \textbf{8.54}} & 47.96 & {\color[HTML]{009901} \textbf{4.19}} & 73.85 & 23.76 & 0.6749 & 0.3536 & 42.60 & 51.22 & 108.67 \\
    Difface (s=100) & 176 & 0.20 & {\color[HTML]{FE0000} \textbf{37.44}} & {\color[HTML]{3531FF} \textbf{4.05}} & 69.34 & 24.83 & 0.6872 & 0.3932 & 46.04 & {\color[HTML]{3531FF} \textbf{45.34}} & {\color[HTML]{FE0000} \textbf{100.78}} \\
    DiffBIR (s=50) & 1667 & 0.07 & 56.61 & 6.16 & {\color[HTML]{FE0000} \textbf{76.51}} & 25.23 & 0.6556 & 0.3839 & 35.24 & {\color[HTML]{FE0000} \textbf{42.30}} & 108.99 \\
    ResShift (s=4)& 195 & 4.26 & 46.95 & {\color[HTML]{000000} 4.28} & 72.85 & {\color[HTML]{009901} \textbf{25.75}} & {\color[HTML]{3531FF} \textbf{0.7048}} & {\color[HTML]{3531FF} \textbf{0.3437}} & {\color[HTML]{3531FF} \textbf{33.82}} & 53.85 & 110.06 \\
    PMRF (s=25)& 176 & 0.63 & {\color[HTML]{3531FF} \textbf{38.52}} & {\color[HTML]{FE0000} \textbf{3.78}} & 71.47 & {\color[HTML]{FE0000} \textbf{26.25}} & {\color[HTML]{FE0000} \textbf{0.7095}} & {\color[HTML]{009901} \textbf{0.3465}} & {\color[HTML]{FE0000} \textbf{30.83}} & 51.82 & {\color[HTML]{3531FF} \textbf{104.72}} \\ \hline
    \textbf{ELIR (Ours)} & {\color[HTML]{FE0000} \textbf{37}} & {\color[HTML]{3531FF} \textbf{19.51}} & {\color[HTML]{009901} \textbf{41.96}} & 4.33 & 70.52 & {\color[HTML]{3531FF} \textbf{25.85}} & 0.7009 & 0.3748 & {\color[HTML]{009901} \textbf{34.38}} & 53.73 & {\color[HTML]{009901} \textbf{106.57}} \\ \bottomrule
    \end{tabular}}
(a) \\ \vspace{0.2cm}
    \resizebox{\textwidth}{!}{%
    \begin{tabular}{lccccccccc}
    \toprule
     & \multicolumn{2}{c}{} & \multicolumn{5}{c}{ImageNet-Validation} & \multicolumn{2}{c}{RealSet80} \\ \cmidrule(l){4-8} \cmidrule(l){9-10}
     & \multicolumn{2}{c}{\multirow{-2}{*}{Efficiency}} & \multicolumn{2}{c}{Perceptual Quality} & \multicolumn{3}{c}{Distortion} & \multicolumn{2}{c}{Perceptual Quality} \\ \cmidrule(l){2-3} \cmidrule(l){4-5} \cmidrule(l){6-8} \cmidrule(l){9-10}
    \multirow{-3}{*}{Model} & \#Params{[}M{]} $\downarrow$ & FPS $\uparrow$ & \multicolumn{1}{l}{NIQE $\downarrow$} & \multicolumn{1}{l}{CLIPIQA $\uparrow$} & PSNR $\uparrow$ & SSIM $\uparrow$ & LPIPS $\downarrow$ & NIQE $\downarrow$ & CLIPIQA $\uparrow$ \\ \midrule\midrule
    BSRGAN  & {\color[HTML]{FE0000} \textbf{16.7}} & {\color[HTML]{3531FF} \textbf{45.85}} & 6.08 & 0.5763 & 22.46 & 0.6298 & 0.3680 & 4.40 & {\color[HTML]{009901} \textbf{0.6162}} \\
    RealESRGAN  & {\color[HTML]{FE0000} \textbf{16.7}} & {\color[HTML]{3531FF} \textbf{45.85}} & {\color[HTML]{009901} \textbf{6.07}} & 0.5306 & 22.26 & 0.6346 & 0.3572 & {\color[HTML]{3531FF} \textbf{4.19}} & 0.6004 \\
    SwinIR-GAN  & 28 & 19.18 & {\color[HTML]{3531FF} \textbf{5.92}} & 0.5532 & 22.11 & {\color[HTML]{009901} \textbf{0.6372}} & 0.3433 & {\color[HTML]{009901} \textbf{4.24}} & 0.5876 \\
    DiffBIR (s=50)  & 1683 & 0.07 & 9.88 & {\color[HTML]{FE0000} \textbf{0.7877}} & 21.41 & 0.5624 & 0.3756 & 6.39 & {\color[HTML]{3531FF} \textbf{0.6308}} \\
    DiffIR (s=4) & {\color[HTML]{009901} \textbf{25}} & {\color[HTML]{009901} \textbf{23.24}} & 6.08 & 0.5440 & {\color[HTML]{3166FF} \textbf{22.88}} & {\color[HTML]{3166FF} \textbf{0.6511}} & {\color[HTML]{3166FF} \textbf{0.3245}} & 4.67 & 0.5658 \\
    Resshift (s=4) & 174 & 4.95 & 7.24 & 0.5941 & {\color[HTML]{FE0000} \textbf{23.13}} & {\color[HTML]{FE0000} \textbf{0.6607}} & {\color[HTML]{FE0000} \textbf{0.2993}} & 5.34 & 0.6127 \\
    SinSR (s=1) & 174 & 19.80 & 6.29 & {\color[HTML]{3166FF} \textbf{0.6091}} & 22.78 & 0.6367 & {\color[HTML]{009901} \textbf{0.3322}} & 5.44 & {\color[HTML]{FE0000} \textbf{0.7100}} \\ \hline
    \textbf{ELIR (Ours)} & {\color[HTML]{3166FF} \textbf{18}} & {\color[HTML]{FE0000} \textbf{52.20}} & {\color[HTML]{FE0000} \textbf{5.27}} & {\color[HTML]{009901} \textbf{0.6041}} & {\color[HTML]{009901} \textbf{22.81}} & 0.6335 & 0.3743 & {\color[HTML]{FE0000} \textbf{4.17}} & 0.6153 \\ \bottomrule
    \end{tabular}}
(b) 
\caption{\textbf{Quantitative Comparison}. Comparison between \name{} and baseline models for (a) BFR and (b) BSR. For iterative methods, we indicate the number of sampling steps by 's' as reported by the authors. \textcolor{red}{Red}, \textcolor{blue}{blue}, and {\color[HTML]{009901} green} indicate the best, the second best, and the third best scores, respectively.}

\label{table:bfr_bsr}
\end{table*}

\section{Experiments}
In this section, we present experiments for the following tasks: blind face restoration (BFR), super-resolution, denoising, inpainting, and blind super-resolution (BSR). The model is trained using the AdamW \cite{loshchilovdecoupled} optimizer. During training, we only use random horizontal flips for data augmentation. We use an exponential moving average (EMA) with a decay of 0.999. The final EMA weights are then used in all evaluations. During inference, we set $M=K$ for Euler steps. The performance evaluation metrics are computed using \citet{pyiqa}, and we report the number of parameters and FPS. FPS is evaluated by injecting images into an NVIDIA GeForce RTX 2080 Ti and recording its process time. Model settings ($K$, \#Params, $\sigma_s$, etc.) were chosen to provide a suitable trade-off between efficiency and balancing distortion and perception quality. The training hyperparameters are provided in the Appx.~\ref{apx:hp}.

\subsection{Implementation Details}\label{impl}
\paragraph{Face Restoration.} We train our model for each task on the FFHQ \cite{karras2019style} dataset, which contains 70k high-quality images. We report FID (vs FFHQ) \cite{heusel2017gans}, NIQE \cite{mittal2012making}, and MUSIQ \cite{ke2021musiq} for perception metrics and PSNR, SSIM \cite{wang2004image}, LPIPS \cite{zhang2018perceptual}, and IDS for distortion metrics. Note that IDS (Identity Score) serves as a quantifier for the identity between the restored images and their ground truths, by gauging the embedding angle of ArcFace \cite{deng2019arcface}. For BFR, the training process is conducted on $512\times512$ resolution with a first-order degradation model to synthesize LQ images. The degradation \cite{zhang2021designing} is approximated by $\vectorsym{y}=\{[(\vectorsym{x} \circledast k_\sigma)\downarrow r+n_\delta]_{\mathbf{JPEG}_{Q}}\}\uparrow r$,
where $\circledast$ denotes convolution, $k_\sigma$ is a Gaussian blur kernel of size $41\times41$ with variance $\sigma^2$, $\downarrow r$ and $\uparrow r$ are down-sampling and up-sampling by a factor $r$, respectively. $n_\delta$ is Gaussian noise with variance $\delta^2$ and $[\cdot]_{\mathbf{JPEG}_{Q}}$ is JPEG compression-decompression with quality factor Q. We choose $\sigma$,$r$,$\delta$,$Q$ uniformly from [0.1, 15], [0.8, 32], [0, 20], and [30, 100], respectively. We set $\sigma_s=0.1$ and the consistency loss is applied with multi-segments of $K=5$. We evaluate our method on the synthetic CelebA-Test \cite{liu2015faceattributes} and on In-The-Wild Face datasets: LFW-Test \cite{LFWTech} and CelebAdult \cite{wang2021towards}. CelebA-Test consists of 3000 pairs of low and high-quality face images taken from CelebA and degraded by \citet{wang2021towards}. For face restoration tasks: super-resolution ($\times$8), denoising, and inpainting (randomly masking 90\% of the pixels), the training process is similar to PMRF \cite{ohayon2024posterior}. We employ a 256$\times$256 resolution and utilize the same degradation model and $\sigma_s$. The consistency loss is applied with multi-segments of $K=3$. We tested our method on the synthetic CelebA-Test, where the same training degradations were used in the evaluation.\\
\paragraph{Image Restoration.} We train our model on the general-content ImageNet \cite{deng2009imagenet} dataset for BSR. Similar to the training process outlined in ResShift \cite{10681246}, we employ a 256$\times$256 resolution and utilize the same degradation model, where the LQ images are $\times$4 downscaled and degraded using the pipeline of RealESRGAN \cite{wang2021realesrgan}. The downscaled images are first ×4 bicubic upscaled before feeding them into the model. We set $\sigma_s=0.04$, and the consistency loss is applied with multi-segments of $K=3$. We test our method on synthetic ImageNet-Validation, consisting of 3000 pairs of LQ and HQ images degraded by ResShift \cite{10681246}, and on the real-world dataset RealSet80 \cite{10681246}. We report NIQE and CLIPIQA \cite{wang2023exploring} for perception metrics and PSNR, SSIM, and LPIPS for distortion metrics.

\begin{figure*}
\centering
\includegraphics[width=1.0\textwidth]{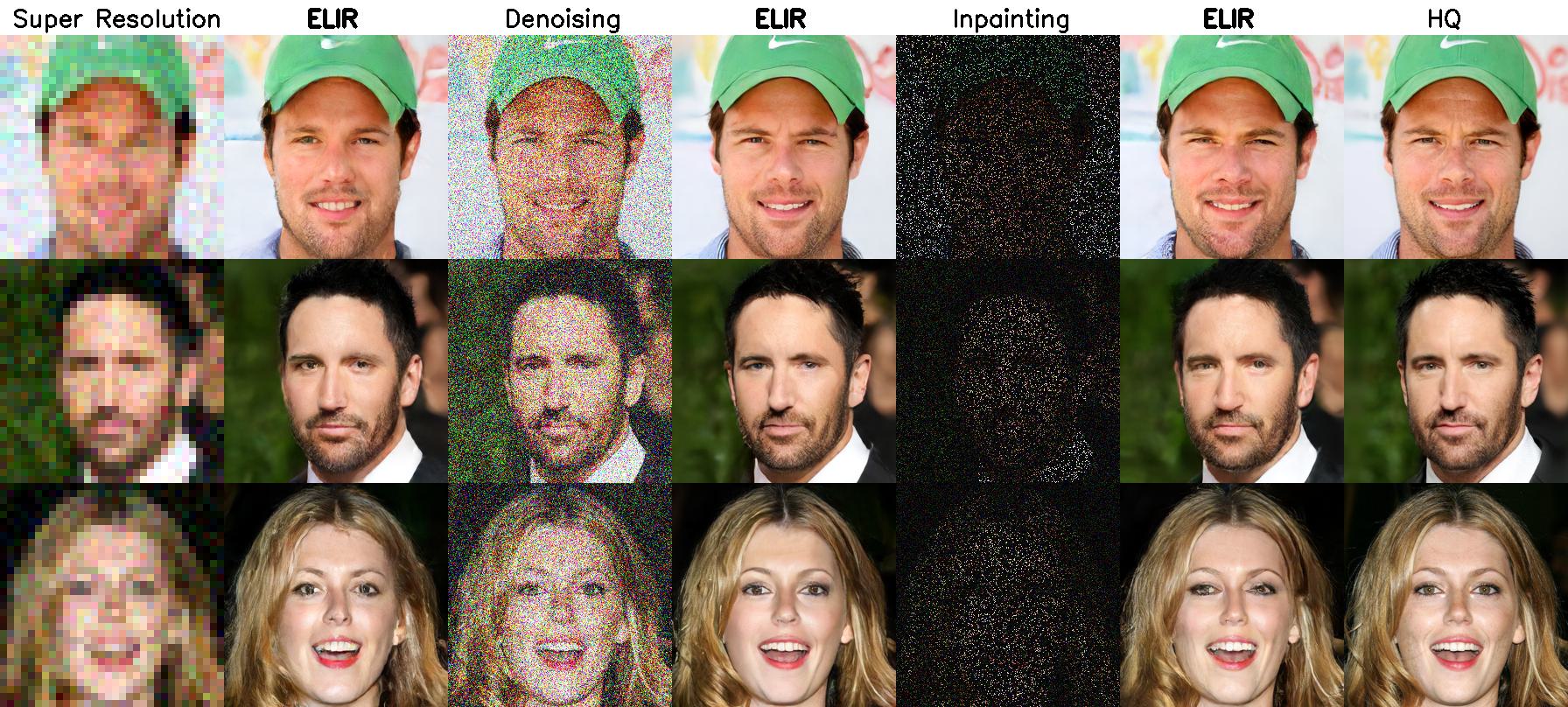}
\caption{\textbf{Face Restoration Visual Results}. Visual results of \name{} for face super resolution ($\times$8), denoising, and inpainting.}
\label{fig:tasks}
\end{figure*}

\subsection{Results}
\paragraph{Face Restoration.} For BFR, we compare our method with the following baseline models: CodeFormer \cite{zhou2022codeformer}, GFPGAN \cite{wang2021gfpgan}, VQFRv2 \cite{gu2022vqfr}, Difface \cite{yue2024difface}, DiffBIR \cite{lin2023diffbir}, ResShift \cite{10681246}, and PMRF \cite{ohayon2024posterior}. In Table~\ref{table:bfr_bsr}(a), we present a comparative evaluation showing that \name{} is competitive with state-of-the-art methods. Our method achieves a notably high PSNR without compromising FID, indicating its ability to balance perception and distortion. Moreover, \name{} has the smallest model size compared to all other methods. In terms of latency, \name{} is much faster compared to diffusion \& flow-based methods. While GAN-based methods exhibit comparable latency to \name{}, they suffer from significant drops in PSNR and IDS, often failing to preserve the person's identity.  In addition, Fig.~\ref{fig:bfr} presents visual results of \name{} compared to baseline methods. With its competitive performance, minimal model size, and fast inference, \name{} is ideally positioned for deployment on resource-constrained devices. Table~\ref{table:tasks} compares \name{} and PMRF \cite{ohayon2024posterior} for super-resolution, denoising, and inpainting, and Fig.~\ref{fig:tasks} presents visual results. Our method achieves competitive performance with PMRF in terms of perceptual quality while exhibiting a slight performance gap in distortion metrics. \name{} demonstrates a $4.6\times$ reduction (only 27M) in model size and a $45\times$ speedup ($\sim$50 FPS) compared to PMRF.

\paragraph{Image Restoration.} We compare our method for BSR with the following baseline methods: BSRGAN \cite{zhang2021designing}, RealESRGAN \cite{wang2021realesrgan}, SwinIR-GAN \cite{liang2021swinir}, DiffIR \cite{xia2023diffir}, DiffBIR \cite{lin2023diffbir}, ResShift \cite{10681246}, and SinSR \cite{wang2024sinsr}. In Table~\ref{table:bfr_bsr}(b), we present a comparative evaluation showing that \name{} is competitive with baseline methods. \name{} is the smallest compared to diffusion-based methods and fastest compared to all methods. Yet, it shows competitive results and effectively balances distortion and perception. Compared to GAN-based methods, \name{} exhibits higher PSNR and FPS. Visual results are provided in Fig.~\ref{fig:bsr}, and additional results in the Appx.~\ref{apx:results}.

\begin{table}
\begin{tabular}{llcc}
\toprule
 & \multicolumn{1}{c}{} &  & \multicolumn{1}{l}{} \\
\multirow{-2}{*}{Task} & \multicolumn{1}{l}{\multirow{-2}{*}{Model}} & \multirow{-2}{*}{FID $\downarrow$} & \multicolumn{1}{l}{\multirow{-2}{*}{PSNR $\uparrow$}} \\ \hline\hline
 & PMRF & 43.24 & 24.33 \\
\multirow{-2}{*}{Super Resolution} & \textbf{ELIR (Ours)} & 44.61 & 24.08 \\ \hline
 & PMRF & 41.42 & 27.87 \\
\multirow{-2}{*}{Denoising} & \textbf{ELIR (Ours)} & 40.26 & 27.21 \\ \hline
 & PMRF & 39.60 & 25.86 \\
\multirow{-2}{*}{Inpainting} & \textbf{ELIR (Ours)} & 39.92 & 25.55 \\ \bottomrule
\end{tabular}
\caption{\textbf{Face Restoration performance.}}
\label{table:tasks}
\end{table}

\subsection{Ablation}
\paragraph{Distortion-Perception trade-off.} This ablation study involves tuning $\beta$ to manage the balance between distortion and perception. Higher $\beta$ values prioritize minimizing distortion over perceptual quality. Our findings in Table~\ref{table:beta} suggest that $\beta=0.001$ offers a suitable compromise, balancing FID and PSNR.
This ablation was conducted on the CelebA-Test dataset for super-resolution, but similar results were found in other tasks.

\begin{table}[H]
\centering
\begin{tabular}{lcc}
\\\toprule
 & \multicolumn{1}{l}{} &  \\
\multirow{-2}{*}{$\beta$} &  \multirow{-2}{*}{FID $\downarrow$} & \multirow{-2}{*}{PSNR $\uparrow$} \\ \midrule\midrule
0.0 & 44.04 &  23.85 \\
0.001 & 44.61 &  24.08 \\ 
0.01 & 46.54 &  24.72\\ \bottomrule
\end{tabular}
\caption{\textbf{Distortion-Perception trade-off}}
\label{table:beta}
\end{table}

\paragraph{Noise Level.} This ablation study investigates the impact of noise level $\sigma_s$ on model performance. Additive noise is vital for learning the complex dynamics of image degradation, enabling the generation of high-quality images. However, careful tuning of $\sigma_s$ is essential; excessive noise can lead to distortion, while insufficient noise may degrade perceptual quality. Table~\ref{table:noise} presents the results for various $\sigma_s$ values. Based on these results, $\sigma_s=0.1$ appears to offer a fair balance between minimizing distortion and maintaining high perceptual quality. This ablation was conducted on the CelebA-Test dataset for super-resolution.

\begin{table}[H]
\centering
\begin{tabular}{lcc}
\\\toprule
 & \multicolumn{1}{l}{} &  \\
\multirow{-2}{*}{$\sigma$} &  \multirow{-2}{*}{FID $\downarrow$} & \multirow{-2}{*}{PSNR $\uparrow$} \\ \midrule\midrule
0.0 & 52.73 &  24.48 \\
0.1 & 44.61 &  24.08 \\ 
0.2 & 41.38 &  23.86 \\ \bottomrule
\end{tabular}
\caption{\textbf{Noise Level}}
\label{table:noise}
\end{table}

\paragraph{Pixel vs Latent.} In this ablation study, we investigate the influence of moving to the latent space on model performance and efficiency. Table~\ref{table:cfm_fm} presents the results of Pixel CFM (w/o encoder-decoder) and Latent FM (w/o consistency) with 25 steps. We can observe that in Pixel CFM, the performance is slightly better, but the efficiency (FPS) is worse, as expected. In addition, Latent FM shows similar performance to ELIR but requires more steps (low FPS). This ablation was conducted on the CelebA-Test dataset for the denoising task. Additional ablations are provided in the Appx.~\ref{apx:ablation}. 

\begin{table}[H]
\centering
\begin{tabular}{lccc}
\\\toprule
 & \multicolumn{1}{l}{} &  \\
\multirow{-2}{*}{Model} &  \multirow{-2}{*}{FID $\downarrow$} & \multirow{-2}{*}{PSNR $\uparrow$} & \multirow{-2}{*}{FPS $\uparrow$}\\ \midrule\midrule
Pixel CFM & 38.50 &  27.54 & 7.32\\
Latent FM & 40.13 &  27.09 & 11.20\\ 
\name{} & 40.26 &  27.21 & 49.26\\ \bottomrule
\end{tabular}
\caption{\textbf{Pixel vs Latent}}
\label{table:cfm_fm}
\end{table}

%% file: files/appendix.tex
\section{Appendix}\label{appendix}
\textbf{Table of Contents:}
\begin{itemize}
    \item Subsection ~\ref{apx:bound}: provides the theoretical proof for the Wasserstein-2 bound.
    \item Subsection ~\ref{apx:fm}: justifies our flow matching source and target distributions.
    \item Subsection ~\ref{apx:nn}: provides a description of the neural network architecture.
    \item Subsection ~\ref{apx:hp}: provides details on the hyperparameters used in the experiments.
    \item Subsection ~\ref{apx:ablation}: contains additional ablations.
    \item Subsection ~\ref{apx:results}: contains additional results.
\end{itemize}

\newpage

\subsection{Wasserstein-2 Bound}\label{apx:bound}

\begin{theorem}
    Let $\vectorsym{x}\in\mathcal{X}\subseteq\mathbb{R}^{d_{x}}$ be a random vector that represents an HQ image, $\hat{\vectorsym{x}}=\mathcal{D}\brackets{\hat{\vectorsym{z}}_1}\in\mathcal{X}\subseteq\mathbb{R}^{d_{x}}$ be a vector that represents the reconstructed image using a random latent variable $\hat{\vectorsym{z}}_1\in\mathcal{Z}\subseteq\mathbb{R}^{d_z}$, $\mathcal{D}:\mathcal{Z}\rightarrow\mathcal{X}$ be a decoder from a latent space to image space, $\vectorsym{z}_1=\mathcal{E}(\vectorsym{x})\in\mathcal{Z}\subseteq\mathbb{R}^{d_z}$ be the latent vector, and $\mathcal{E}:\mathcal{X}\rightarrow\mathcal{Z}$ be an encoder from a image to latent space. To obtain $\hat{\vectorsym{z}}_1$ we solve the following ODE at time $t=1$:  \begin{equation}\label{eq:ode_solve}
        \frac{d\hat{\vectorsym{z}}_t}{dt} = \hat{\vectorsym{v}}(\hat{\vectorsym{z}}_t,t), \quad \hat{\vectorsym{z}}_0=\vectorsym{z}_0
    \end{equation}
    where $\vectorsym{z}_0$ is some predefined source distribution (usually a standard Gaussian distribution), $\vectorsym{z}_1$ is the target distribution, and $\hat{\vectorsym{v}}\brackets{\vectorsym{z}_t,t}$ is the learned velocity field. Define ${\vectorsym{v}}\brackets{\vectorsym{z}_t,t}$ as the velocity field which obtains $\vectorsym{z}_1$ by solving \eqref{eq:ode_solve} and Wasserstein-2 distance between two random variables:
\begin{equation}
    W_2(p_{\hat{\vectorsym{x}}},p_{\vectorsym{x}})\triangleq \brackets{\inf\limits_{\mu\in\Pi\brackets{p_{\vectorsym{x}},p_{\hat{\vectorsym{x}}}}}\int_{\mathcal{X}\times\mathcal{X}}\norm{\x-\hat{\x}}^2d\mu\brackets{\x,\hat{\x}}}^{\frac{1}{2}}
\end{equation}
where $\Pi$ is the set of probability measures of $\mu$ on $\mathcal{X}\times\mathcal{X}$. The encoder-decoder error $\Delta_{\mathcal{E},\mathcal{D}}$ and the vector field error $\Delta_{\vectorsym{v}}$ are defined as:
\begin{align*}
    \Delta_{\mathcal{E},\mathcal{D}}&\triangleq{\expectation{\norm{\mathcal{D}(\mathcal{E}(\vectorsym{x}))-\x}^2}{\x}},\\
    \Delta_{\vectorsym{v}} &\triangleq {\int_{t=0}^{1}\int_{\vectorsym{z}\in\mathcal{Z}}\norm{\vectorsym{v}\brackets{\z,t}-\hat{\vectorsym{v}}\brackets{\z,t}}^2 p_{\z_t}\brackets{\z} d\vectorsym{z}dt}=\int_{t=0}^1 \expectation{\norm{\vectorsym{v}\brackets{\z_t,t}-\vectorsym{v}_{\theta}\brackets{\z_t,t}}^2}{\z_t}dt.
\end{align*}
Assume that $\mathcal{D}$ is a Lipschitz function with constant $L_{\mathcal{D}}$ and that the learned velocity field $\hat{\vectorsym{v}}\brackets{\vectorsym{z}_t,t}$ is continuously differentiable and k-Lipschitz in $\vectorsym{z}_t$ throughout the domain with Lipschitz constant $L_{\hat{\vectorsym{v}}}$. Then, the Wasserstein-2 distance between the HQ image and the reconstructed image is bounded by:
\begin{align}\label{bound2_final}
    W_2(p_{\hat{\vectorsym{x}}},p_{\vectorsym{x}}) &\leq \sqrt{\Delta_{\mathcal{E},\mathcal{D}}} + L_{\mathcal{D}} e^{0.5+L_{\hat{\vectorsym{v}}}}\sqrt{\Delta_{\vectorsym{v}}}=\sqrt{\Delta_{\mathcal{E},\mathcal{D}}} + C\sqrt{\Delta_{\vectorsym{v}}}.
\end{align}
\end{theorem}

\proof{} Let $\tilde{\vectorsym{x}}=\mathcal{D}(\mathcal{E}(\vectorsym{x}))=\x+\delta_{\mathcal{E},\mathcal{D}}(\x)$ be an HQ image with encoder-decoder error where $\delta_{\mathcal{E},\mathcal{D}}(\x)$ is the encoder-decoder error function depending on the sample $\x$. Then, by using the triangle inequality for the Wasserstein-2 metric \cite{panaretos2019statistical}, we have:
\begin{align}\label{bound2}
    W_2(p_{\hat{\vectorsym{x}}},p_{\vectorsym{x}}) &\leq W_2(p_{\tilde{\vectorsym{x}}},p_{\vectorsym{x}}) + W_2(p_{\tilde{\vectorsym{x}}},p_{\hat{\vectorsym{x}}}),
\end{align}
Now we investigate the two terms in \eqref{bound2}. For the first term, we obtain:
\begin{align}\label{bound_first}
    W_2(p_{\tilde{\vectorsym{x}}},p_{\vectorsym{x}})&= \brackets{\inf\limits_{\mu\in\Pi\brackets{p_{\vectorsym{x}},p_{\tilde{\vectorsym{x}}}}}\int_{\mathcal{X}\times\mathcal{X}}\norm{\x-\tilde{\x}}^2d\mu\brackets{\x,\tilde{\x}}}^{\frac{1}{2}}\\ \nonumber
    &=\brackets{\inf\limits_{\mu\in\Pi\brackets{p_{\vectorsym{x}},p_{\tilde{\vectorsym{x}}}}}\int_{\mathcal{X}\times\mathcal{X}}\norm{\tilde{\x}-\tilde{\x}+\delta_{\mathcal{E},\mathcal{D}}\brackets{\x}}^2d\mu\brackets{\x,\tilde{\x}}}^{\frac{1}{2}}\\ \nonumber
    &\leq\sqrt{\expectation{\norm{\delta_{\mathcal{E},\mathcal{D}}\brackets{\x}}^2}{\x}} = \sqrt{\Delta_{\mathcal{E},\mathcal{D}}}.
\end{align}
 The expectation in the final part of \eqref{bound_first} is derived by marginalizing $\tilde{\x}$. As for the second term in \eqref{bound2} we have that:
\begin{align}\label{eq:k_lip_decode}
     W_2(p_{\tilde{\vectorsym{x}}},p_{\hat{\vectorsym{x}}})&=\brackets{\inf\limits_{\mu\in\Pi\brackets{p_{\hat{\vectorsym{x}}},p_{\tilde{\vectorsym{x}}}}}\int_{\mathcal{X}\times\mathcal{X}}\norm{\hat{\x}-\tilde{\x}}^2d\mu\brackets{\hat{\x},\tilde{\x}}}^{\frac{1}{2}}\nonumber\\
     &=\brackets{\inf\limits_{\mu\in\Pi\brackets{p_{\hat{\vectorsym{z}}_1},p_{\vectorsym{z}_1}}}\int_{\mathcal{Z}\times\mathcal{Z}}\norm{\mathcal{D}\brackets{\hat{\vectorsym{z}_1}}-\mathcal{D}\brackets{\vectorsym{z}_1}}^2d\mu\brackets{\hat{\vectorsym{z}}_1,\vectorsym{z}_1}}^{\frac{1}{2}}\nonumber\\
     &\leq L_{\mathcal{D}} \brackets{\inf\limits_{\mu\in\Pi\brackets{p_{\hat{\vectorsym{z}}_1},p_{\vectorsym{z}_1}}}\int_{\mathcal{Z}\times\mathcal{Z}}\norm{{\hat{\vectorsym{z}}_1}-{\vectorsym{z}_1}}^2d\mu\brackets{\hat{\vectorsym{z}}_1,\vectorsym{z}_1}}^{\frac{1}{2}}= L_{\mathcal{D}}\cdot W_2(p_{{\vectorsym{z}}_1},p_{\hat{\vectorsym{z}}_1}).
\end{align}
In the first step, we replace $\x$ with $\mathcal{D}\brackets{\z_1}$, which is a direct result of the expectation form of $W_2$ \cite{panaretos2019statistical} combined with the law of the unconscious statistician (LOUTS), and in the last step, we use the fact that $\mathcal{D}$ is a k-Lipschitz function. Next, we bound the Wasserstein-2 distance of the vector field error using Proposition 3 from \citet{AlbergoV23}, and note that since we add Gaussian noise to $\vectorsym{z}_0$ it is supported on all $\mathbb{R}^{d_z}$. 
As shown in \citet{AlbergoV23}, the Wasserstein-2 distance between $\vectorsym{z}_1$ and $\hat{\vectorsym{z}}_1$ can be bounded by the error of the learned velocity field $\hat{\vectorsym{v}}\brackets{\vectorsym{z}_t,t}$ with the true velocity $\vectorsym{v}\brackets{\vectorsym{z}_t,t}$. Specifically, using the assumption that $\hat{\vectorsym{v}}\brackets{\vectorsym{z}_t,t}$ is continuously differentiable and k-Lipschitz in $\vectorsym{z}_t$ on the entire domain with Lipschitz constant $L_{\hat{\vectorsym{v}}}$, we have the following:
\begin{equation}\label{bound_second}
    W_2(p_{{\vectorsym{z}_1}},p_{\hat{\vectorsym{z}}_1}) \leq e^{0.5+L_{\hat{\vectorsym{v}}}}\sqrt{\int_{t=0}^{1}\int_{\vectorsym{z}\in\mathcal{Z}}\norm{\vectorsym{v}\brackets{\z,t}-\hat{\vectorsym{v}}\brackets{\z,t}}^2 p_{\z_t}\brackets{\z} d\vectorsym{z}dt}.
\end{equation}
Finally combined \eqref{bound2}, \eqref{bound_first}, \eqref{eq:k_lip_decode} and \eqref{bound_second} results in \eqref{bound2_final}.

\subsection{Source and Target distributions}\label{apx:fm}
While \citet{lipman2023flow} treated the source distribution as a known prior (e.g., a standard Gaussian), Rectified Flow \cite{liu2023flow} and OT-CFM \cite{tong2024improving} showed that matching flows can be trained between any source and target distributions. Specifically, OT-CFM (Optimal transport CFM) generalized this to distribution $p(\vectorsym{z}_0, \vectorsym{z}_1)$ where $\vectorsym{z}_0$ and $\vectorsym{z}_1$ are dependent, as in our paired dataset. Unlike scenarios where $\vectorsym{z}_0$ and $\vectorsym{z}_1$ are sampled independently, here, they are sampled jointly. In our work, $\vectorsym{z}_0$ and $\vectorsym{z}_1$, the MMSE output and the latent representation of its corresponding high-quality image, are sampled jointly.

\subsection{Neural Network Architecture}\label{apx:nn}
To achieve a lightweight and efficient model, we utilize Tiny AutoEncoder \cite{von-platen-etal-2022-diffusers}, a pre-trained tiny CNN version of Stable Diffusion VAE \cite{esser2024scaling}. Tiny AutoEncoder allows us to compress the image by a factor of 12, e.g., $CHW=(3,512,512)$ into $CHW=(16,64,64)$, with only 1.2M parameters for each encoder and decoder. Given the model size and latency constraints, we restrict our architecture to convolutional layers only, eschewing transformers' global attention mechanisms. Linear operations such as convolution can be modeled as matrix multiplication with a little overhead. As a result, these operations are highly optimized on most hardware accelerators to avoid quadratic computing complexity. Although Windows attention techniques \cite{liang2021swinir,crowson2024scalable} can be theoretically implemented with linear time complexity, practical implementation often involves data manipulation operations, including reshaping and indexing, which remain crucial considerations for efficient implementation on resource-constrained devices. Alternatively, in our method, we use only convolution layers.

\subsubsection{Coarse Estimator}
The coarse estimator consists of 3 cascaded RRDB blocks \cite{wang2018esrgan} with 96 channels each. We replace the Leaky ReLU activation of the original RRDB with SiLU. The cascade is implemented with a skip connection. 

\subsubsection{U-Net}
For implementing the vector field, we use U-Net \cite{ronneberger2015u}. U-Net is an architecture with special skip connections. These skip connections help transfer lower-level information from shallow to deeper layers. Since the shallower layers often contain low-level information, these skip connections help improve the result of image restoration. Our U-Net consists of convolution layers only. It has 3 levels with channel widths of $(128,256,512)$ and depths of $(1,2,4)$. We add a first and last convolution to align the channels of the latent tensor shape. Our basic convolution layer has a $3\times3$ kernel, and all activation functions are chosen to be SiLU. During training, we utilize collapsible linear blocks \cite{bhardwaj2022collapsible} by adding $1\times1$ convolution after each $3\times3$ convolution layer and expanding the hidden channel width by $\times$4. These two linear operations are then collapsed to a single $3\times3$ convolution layer before inference.

\subsection{Hyper-parameters}\label{apx:hp}
\begin{table}[H]
\centering
\footnotesize
\begin{tabular}{lccc}
\hline
 &  &  & \multicolumn{1}{l}{} \\
\multirow{-2}{*}{Hyper-parameter} & \multirow{-2}{*}{Blind Face Restoration} & \multirow{-2}{*}{Face Restoration tasks} & \multicolumn{1}{l}{\multirow{-2}{*}{Blind Super Resolution}} \\ \midrule
Vector-Field parameters & 29M &  19M & 10M \\
Coarse Estimator parameters & 5.5M & 5.5M & 5.5M \\
Encoder-Decoder parameters & 2.4M & 2.4M & 2.4M \\
Euler steps ($M$) & 5 & 3 & 3 \\
CFM segments ($K$) & 5 & 3 & 3 \\
CFM $\Delta t$ & 0.05 & 0.05 & 0.2 \\
CFM $\alpha$ & 0.001 & 0.001 & 0.001 \\
$\beta$ & 0.001 & 0.001 & 0.001 \\
$\sigma_{min}$ & $10^{-5}$ & $10^{-5}$ & $10^{-5}$ \\
Training epochs & 400 &  250 & 50 \\
Batch size & 64 & 128 & 128 \\
Image dimension & 3$\times$512$\times$512 & 3$\times$256$\times$256 & 3$\times$256$\times$256 \\
Latent dimension & 16$\times$64$\times$64 & 16$\times$32$\times$32 & 16$\times$32$\times$32 \\
Training hardware & 4$\times$ H100 80GB & 4$\times$ A100 40GB & 4$\times$ H100 80GB \\
Training time & 2.5 days & 1 days & 2 days \\
Optimizer & AdamW & AdamW & AdamW \\
Learning rate & $10^{-4}$ & $2\cdot10^{-4}$ & $2\cdot10^{-4}$ \\
AdamW betas & (0.9,0.999) & (0.9,0.999) & (0.9,0.999) \\
AdamW eps & $10^{-8}$ & $10^{-8}$ & $10^{-8}$ \\
Weight decay & 0.02 & 0.02 & 0.02 \\
EMA decay & 0.999 & 0.999 & 0.999 \\ \bottomrule
\end{tabular}
\vspace{0.3cm}
\caption{\textbf{Hyper-parameters.} Training hyper-parameters for face and image restoration.}
\label{hyperparams}
\end{table}

\subsection{Additional Ablations}\label{apx:ablation}

\paragraph{Effectiveness of the coarse estimator and LCFM.} This ablation highlights the crucial role of the coarse estimator and LCFM. As shown in Table~\ref{table:coarse}, removing the coarse estimator component leads to a substantial performance drop in PSNR and FID. Removing LCFM leads to a significant drop in perceptual quality. Experiments are conducted for blind face restoration and super-resolution on CelebA-Test. 
\begin{table}[H]
\centering
\resizebox{\textwidth}{!}{%
\begin{tabular}{lccccccccc}
\toprule
 & \multicolumn{1}{l}{} & \multicolumn{1}{l}{} & \multicolumn{3}{c}{Perceptual Quality} & \multicolumn{3}{c}{Distortion} \\ \cmidrule(l){4-6} \cmidrule(l){7-10}
\multirow{-2}{*}{Task} & \multicolumn{1}{l}{\multirow{-2}{*}{\begin{tabular}[c]{@{}l@{}}Coarse\\  Estimator\end{tabular}}} & \multicolumn{1}{l}{\multirow{-2}{*}{LCFM}} & FID $\downarrow$ & \multicolumn{1}{l}{NIQE $\downarrow$} & MUSIQ $\uparrow$ & PSNR $\uparrow$ & SSIM $\uparrow$ & LPIPS $\downarrow$ & IDS $\downarrow$ \\ \midrule\midrule
 & \cmark & \xmark &  76.33 & 7.48 & 47.50 & 26.19 & 0.7080 & 0.4457 & 35.16\\
 & \xmark & \cmark &  44.35 & 4.54 & 66.00 & 25.57 & 0.6985 & 0.3919 & 36.28\\
\multirow{-3}{*}{Blind Face Restoration} & \cmark & \cmark & 41.96 &  4.33 & 70.61 &  25.85 & 0.7009 &  0.3748 & 34.38\\ \midrule
 & \cmark & \xmark &  81.05 & 7.86 & 51.19 & 24.69 & 0.6818 & 0.3639 & 53.35\\
 & \xmark & \cmark & 47.55 & 4.93 & 62.81 & 23.69 & 0.6495 & 0.3401 & 54.13\\
\multirow{-3}{*}{Super Resolution} & \cmark & \cmark & 44.61 &  5.07 & 63.25 & 24.08 & 0.6631 &  0.3252 & 51.42 \\ \bottomrule
\end{tabular}}
\caption{\textbf{Coarse estimator and LCFM}}
\label{table:coarse}
\end{table}

\paragraph{Effectiveness of trainable encoder.} This ablation study demonstrates the importance of fine-tuning the encoder. Given that the encoder was initially trained on HQ images, it struggles to represent the LQ images encountered in various tasks. This limitation is evident in Table~\ref{table:encoders}, where fixed encoders exhibit significantly lower performance, with PSNR values 1.5-2 dB lower and FID scores 1-4 points higher compared to trainable encoders. The experiments were conducted for denoising and inpainting on the CelebA-Test dataset \emph{w/} and \emph{w/o} training the encoder.
\begin{table}[H]
\centering
\begin{tabular}{lcccccccc}
\toprule
 & \multicolumn{1}{l}{} & \multicolumn{3}{c}{Perceptual Quality} & \multicolumn{4}{c}{Distortion} \\ \cmidrule(l){3-5} \cmidrule(l){6-9}
\multirow{-2}{*}{Task} & \multicolumn{1}{l}{\multirow{-2}{*}{\begin{tabular}[c]{@{}l@{}}Trainable\\  Encoder\end{tabular}}} & FID $\downarrow$ & \multicolumn{1}{l}{NIQE $\downarrow$} & MUSIQ $\uparrow$ & PSNR $\uparrow$ & SSIM $\uparrow$ & LPIPS $\downarrow$ & IDS $\downarrow$\\ \midrule\midrule
 & \xmark & 41.55 & 5.04 & 64.39 & 26.55 & 0.7557 & 0.2718 & 38.99\\
\multirow{-2}{*}{Denoising} & \cmark & 40.26 &  5.00 & 65.88 &  27.21 & 0.7748 &  0.2541 & 36.05\\ \midrule
 & \xmark & 43.91 & 5.05 & 62.75 & 23.55 & 0.6658 & 0.3368 & 55.21\\
\multirow{-2}{*}{Inpainting} & \cmark & 39.92 &  4.87 & 65.05 & 25.55 & 0.7330 &  0.2786 & 40.69\\ \bottomrule
\end{tabular}
\caption{\textbf{Trainable encoders}}
\label{table:encoders}
\end{table}

\paragraph{Efficiency of LCFM.} Fig.~\ref{fig:eff}(a) compares the performance of Latent FM and LCFM by plotting PSNR and FID for varying NFEs. Both methods exhibit a similar trend: PSNR decreases while FID improves with increasing NFE, reflecting the expected distortion-perception trade-off. While FM requires 25 NFEs to reach a comparable FID, LCFM achieves the same FID with only 3 NFEs, highlighting LCFM's superior efficiency.
\begin{figure}[H]
\begin{tabular}{cc}
\includesvg[width=0.45\textwidth]{images/fm_cfm.svg}&
\includesvg[width=0.5\textwidth]{images/size_ablation.svg}\\
(a)&(b)
\end{tabular}
\caption{\textbf{Efficiency Ablation.} (a) PSNR and FID vs NFEs. (b) Model Size.}
\label{fig:eff}
\end{figure}

\paragraph{Model Size Ablation.} Table~\ref{table:size} presents different model sizes of \name{} for super-resolution on CelebA-Test. We vary the vector field size while keeping the size of the coarse estimator constant. Our results indicate a diminishing return in FID improvement beyond 27M parameters, as can be shown in Fig.~\ref{fig:eff}(b). 
\begin{table}[H]
\centering
\begin{tabular}{ccccccccc}
\toprule
 & \multicolumn{3}{c}{Percepual Quality} & \multicolumn{4}{c}{Distortion} &  \\ \cmidrule(l){2-4} \cmidrule(l){5-8} 
\multirow{-2}{*}{\#Params {[}M{]}} & FID $\downarrow$ & NIQE $\downarrow$ & MUSIQ $\uparrow$ & PSNR $\uparrow$ & SSIM $\uparrow$ & LPIPS $\downarrow$ & IDS $\downarrow$ & \multirow{-2}{*}{FPS $\uparrow$} \\ \midrule\midrule
13 & 50.84 & 5.00 & 62.37 & 24.05 & 0.6626 & 0.3309 & 51.62 & 50.55 \\
19 & 46.09 & 5.08 & 62.81 & 24.09 & 0.6642 & 0.3267 & 51.54 & 50.05 \\
27 & 44.61 & 5.07 & 63.25 & 24.08 &  0.6631 &  0.3252 & 51.42 & 49.26 \\
37 & 44.59 & 5.07 & 63.28 & 24.09 & 0.6634 & 0.3251 & 51.42 & 42.94 \\ \bottomrule
\end{tabular}
\caption{\textbf{Model Size}}
\label{table:size}
\end{table}

\paragraph{Model Latency Ablation.} Table~\ref{table:latency} presents an ablation study for model latency. Here, we vary the multi-segment value $K$ while maintaining a fixed model size. Each model was trained with a fixed size of 27M parameters. We note that the FPS decreases as $K$ increases. Our findings suggest that $K=3$ provides a suitable trade-off between FID and FPS. This ablation was conducted on the CelebA-Test dataset for the super-resolution, but similar results were found in other face restoration tasks.
\begin{table}[H]
\centering
\begin{tabular}{ccccccccc}
\toprule
 & \multicolumn{3}{c}{Perceptual Quality} & \multicolumn{4}{c}{Distortion} &  \\ \cmidrule(l){2-4} \cmidrule(l){5-8} 
\multirow{-2}{*}{K} & \multicolumn{1}{l}{FID $\downarrow$} & \multicolumn{1}{l}{NIQE $\downarrow$} & MUSIQ $\uparrow$ & PSNR $\uparrow$ & SSIM $\uparrow$ & LPIPS $\downarrow$ & IDS $\downarrow$ & \multirow{-2}{*}{FPS($\uparrow$)} \\ \midrule\midrule
1 & 56.27 & 4.23 & 65.18 & 23.23 & 0.6344 & 0.3609 & 51.17 & 70.30 \\
3 & 44.61 &  5.07 & 63.25 & 24.08 &  0.6631 &  0.3252 & 51.42 & 49.26 \\
5 & 43.97 & 5.11 & 62.95 & 24.18 & 0.6664 & 0.3248 & 51.35 & 36.35 \\ \bottomrule
\end{tabular}
\caption{\textbf{Latency}}
\label{table:latency}
\end{table}

\paragraph{Time Interval Ablation.} In this ablation study, we investigate the influence of the time interval ($\Delta t$) on model performance. Table~\ref{table:delta_t} presents the results of several $\Delta t$ values. Reducing $\Delta t$ is expected to enhance FID scores, however, it may also lead to an increase in distortion metrics. This study aims to identify the $\Delta t$ value that minimizes distortion while maintaining a high level of perceptual quality. According to the results, $\Delta t=0.05$ offers a favorable balance between FID and PSNR. This ablation was conducted on the CelebA-Test dataset for the denoising, but similar results were found in other face restoration tasks.
\begin{table}[H]
\centering
\begin{tabular}{cccccccc}
\toprule
 & \multicolumn{3}{c}{Perceptual Quality} & \multicolumn{4}{c}{Distortion} \\ \cmidrule(l){2-4} \cmidrule(l){5-8}
\multirow{-2}{*}{$\Delta t$} & FID($\downarrow$) & \multicolumn{1}{l}{NIQE $\downarrow$} & MUSIC $\uparrow$ & PSNR $\uparrow$ & SSIM $\uparrow$ & LPIPS $\downarrow$ & IDS $\downarrow$\\ \midrule\midrule
0.01 & 39.92 &  4.85 & 66.36 & 27.10 & 0.7714 & 0.2587 & 36.28 \\
0.05 & 40.26 & 5.00 & 65.88 & 27.21 & 0.7748 & 0.2541 & 36.05 \\
0.1 & 41.63 & 5.26 & 65.43 & 27.25 & 0.7764 & 0.2521 & 36.17 \\ \bottomrule
\end{tabular}
\caption{\textbf{Time Interval}}
\label{table:delta_t}
\end{table}

\subsection{Additional Results}\label{apx:results}

\begin{figure}[H]
\centering
\includegraphics[scale=0.25]{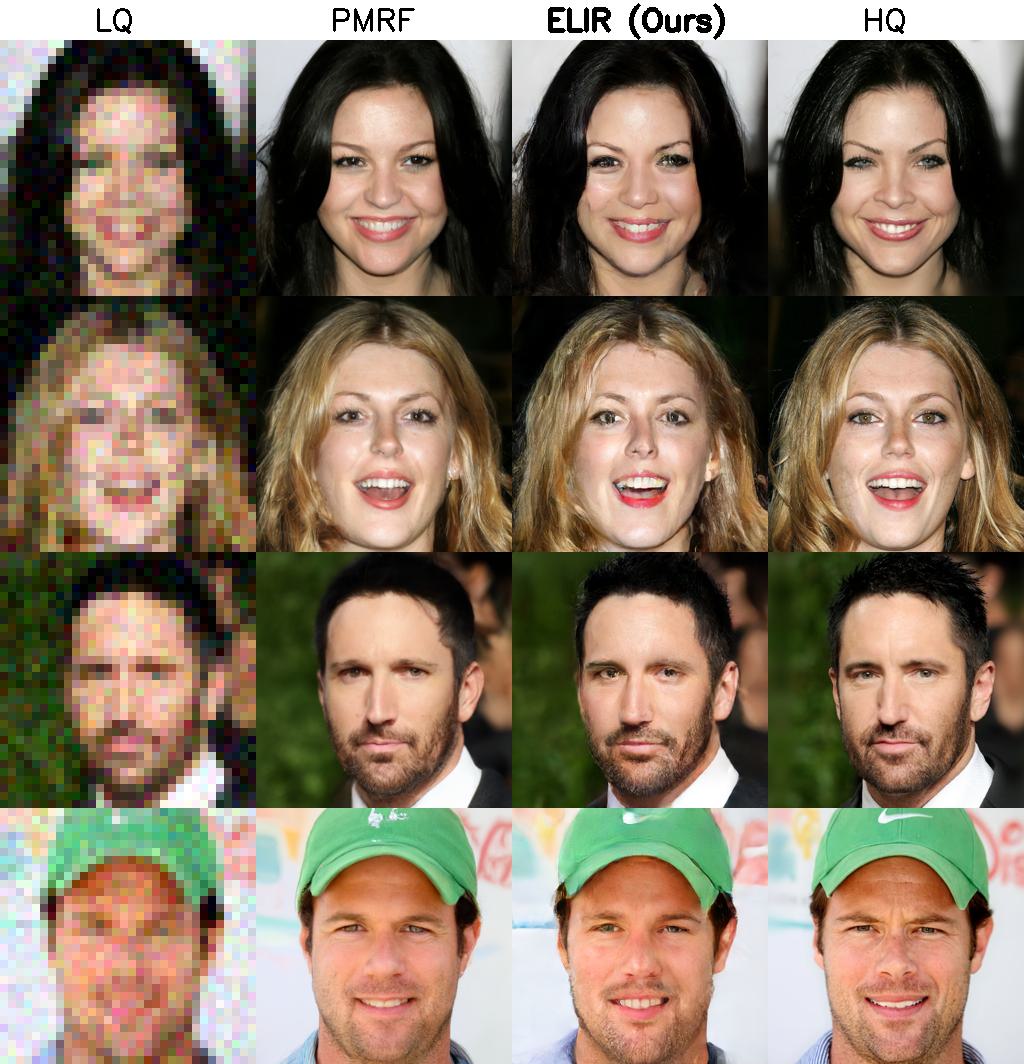}
\caption{\textbf{Visual examples for Super Resolution ($\times$8).} Comparisons between \name{} and PMRF \cite{ohayon2024posterior} sampled from CelebA-Test.}
\label{color_figure}
\end{figure}

\begin{figure}[H]
\centering
\includegraphics[scale=0.25]{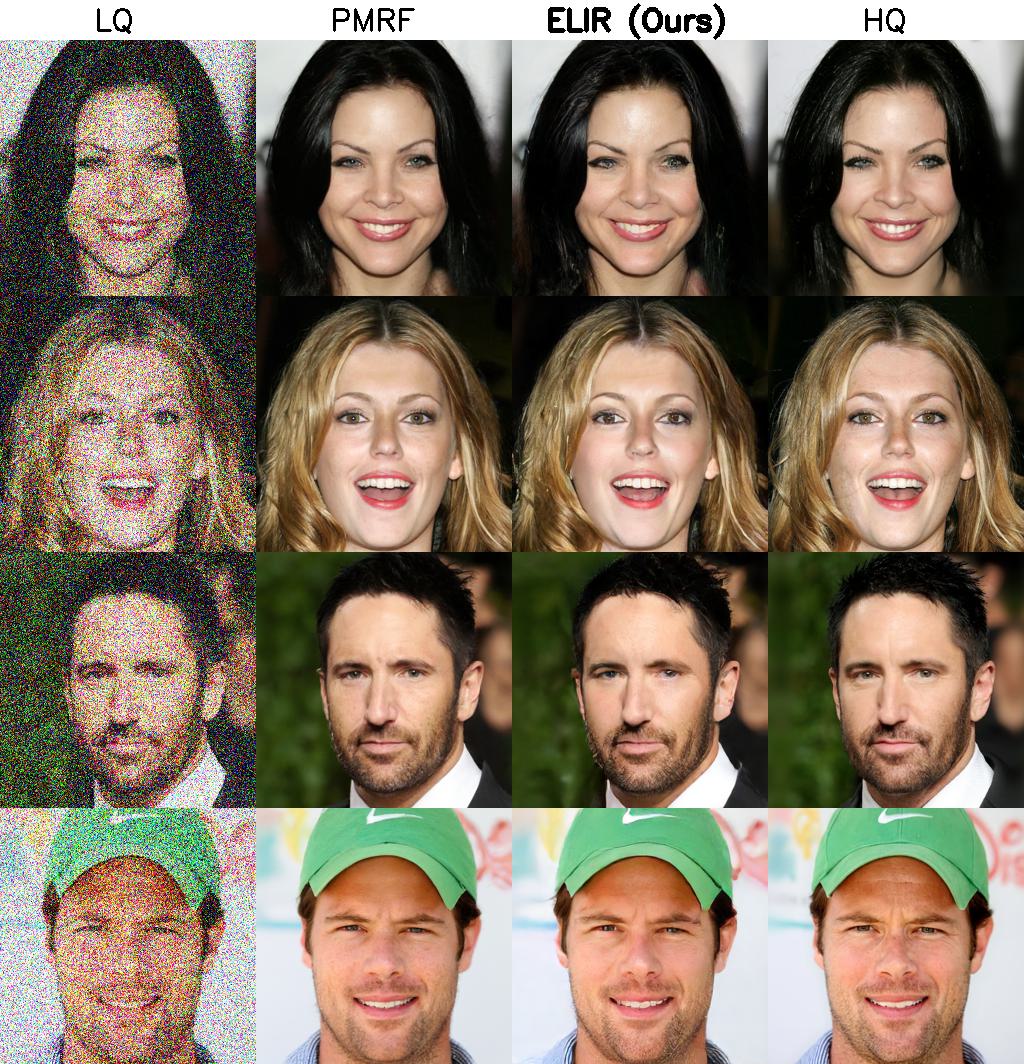}
\caption{\textbf{Visual examples for Denoising.} Comparisons between \name{} and PMRF \cite{ohayon2024posterior} sampled from CelebA-Test.}
\label{id_figure}
\end{figure}

\begin{figure}[H]
\centering
\includegraphics[scale=0.25]{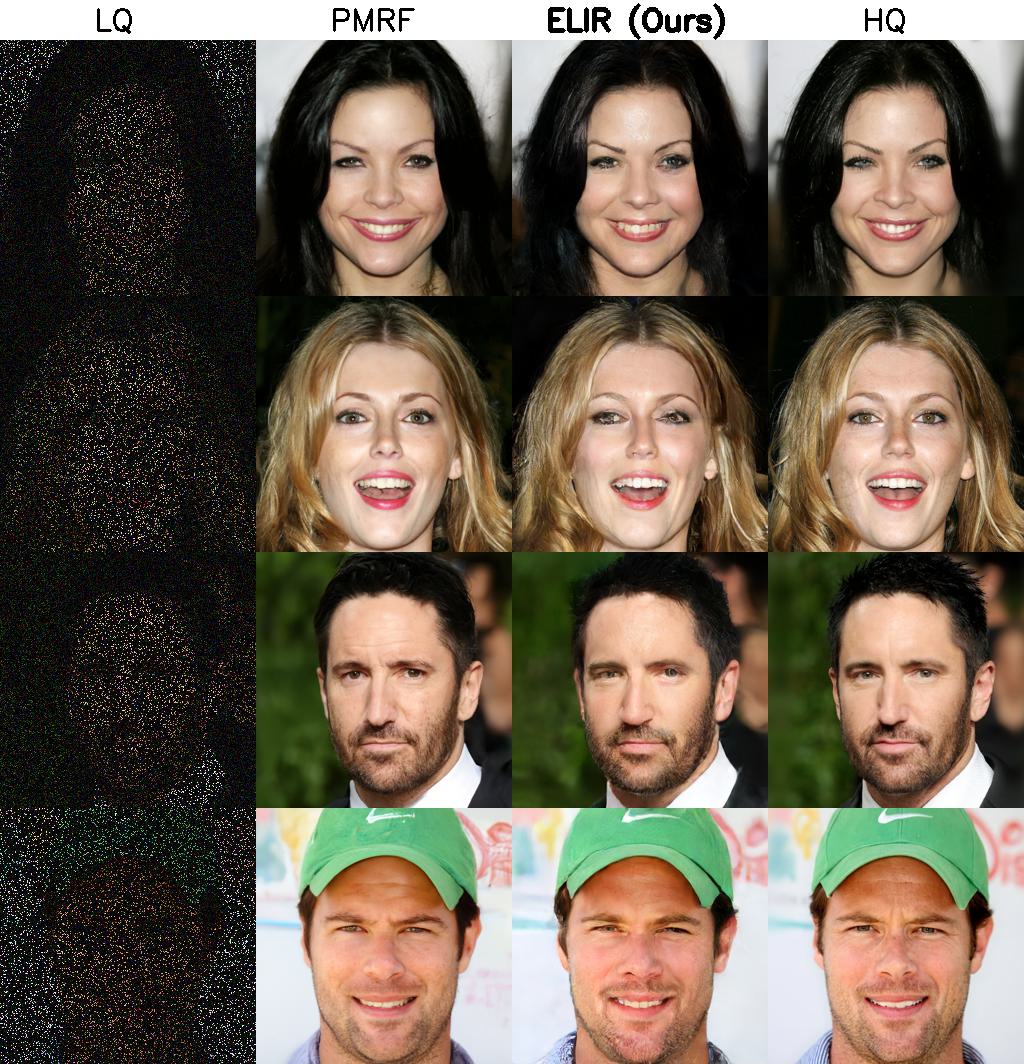}
\caption{\textbf{Visual examples for Inpainting.} Comparisons between \name{} and PMRF \cite{ohayon2024posterior} sampled from CelebA-Test.}
\label{ip_figure}
\end{figure}

\begin{figure}[H]
\centering
\includegraphics[scale=0.1]{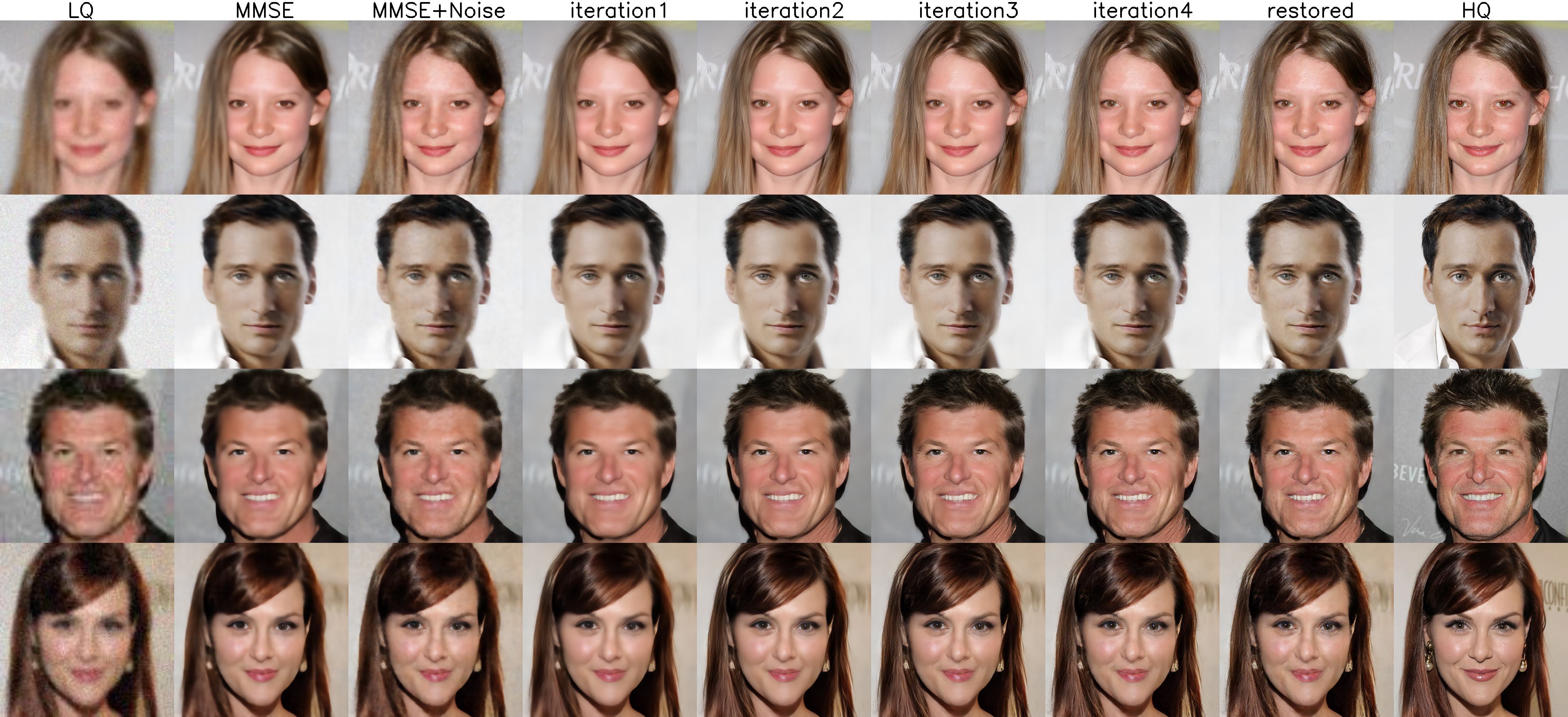}
\caption{\textbf{Visual steps of \name{}.} Illustrating the restoration steps, visualizing the process from LQ images to visually appealing results. The images are sampled from CelebA-Test for blind face restoration.}
\label{fig:flow_bfr}
\end{figure}

\paragraph{LCFM trajectories.} LCFM improves the flow straightness by enforcing consistency within the velocity field, which reduces discretization errors. Fig.~\ref{fid:trajs} illustrates the ``straitness'' of the trajectories in the latent space. However, when these trajectories are projected back to the pixel space, this property is not preserved due to the decoder's non-linearity.
\begin{figure}[H]
\begin{tabular}{cc}
\includesvg[scale=0.5]{images/latent_traj.svg}&
\includesvg[scale=0.5]{images/pixel_traj.svg}\\
(a)&(b)
\end{tabular}
    \caption{\textbf{LCFM trajectories.} (a) visualizes CFM trajectories in latent space, connecting flow from the source (p0) to the target point (p1). These trajectories exhibit ``straight'' flows along two latent variables, a consequence of the LCFM operating within the latent space. However, this linearity is not preserved when projected into pixel space due to the decoder's non-linearity, as demonstrated in (b).}
\label{fid:trajs}
\end{figure}